\documentclass[3p,sort&compress
]{elsarticle}

\makeatletter
\def\@author#1{\g@addto@macro\elsauthors{\normalsize%
    \def\baselinestretch{1}%
    \upshape\authorsep#1\unskip\textsuperscript{%
      \ifx\@fnmark\@empty\else\unskip\sep\@fnmark\let\sep=,\fi
      \ifx\@corref\@empty\else\unskip\sep\@corref\let\sep=,\fi
      }%
    \def\authorsep{\unskip,\space}%
    \global\let\@fnmark\@empty
    \global\let\@corref\@empty  
    \global\let\sep\@empty}%
    \@eadauthor={#1}
}

\usepackage[usenames,dvipsnames,pdftex]{xcolor}

\usepackage{acro} 
\usepackage{amsmath,amssymb,amsfonts,mathrsfs}
\usepackage{miller}
\usepackage{multirow}
\usepackage{subeqnarray}
\usepackage[hang]{caption}
\usepackage[labelfont=footnotesize,textfont=footnotesize]{subcaption}
\usepackage{tikz}
\usetikzlibrary{arrows,shapes,fadings,backgrounds,decorations.shapes,patterns}
\usepackage{verbatim}
\usepackage{bm}
\usepackage{diagbox} 
\usepackage{booktabs}
\usepackage[separate-uncertainty = true]{siunitx}
\usepackage[percent]{overpic}
\usepackage[title]{appendix}
\usepackage{graphicx}
\usepackage{algorithm2e}
\usepackage[pdftex,                                     
bookmarksnumbered=true,
colorlinks=true,%
linkcolor=Ref,
citecolor=Ref,
urlcolor=Ref,
pdfpagelabels=false,
plainpages = false,
]{hyperref}%
\usepackage{doi}

\usepackage[capitalise]{cleveref}                                   
\crefname{algocf}{alg.}{algs.}
\Crefname{algocf}{Algorithm}{Algorithms}

\newlength{\diagramsize}
\DeclareGraphicsExtensions{.pdf,.png}

\newcommand{\revision}[1]{\textcolor{Black}{#1}}

\newcommand{\ie}{\textit{i.e.}}
\newcommand{\eg}{\textit{e.g.}}

\newcommand{\norm}[2][]{\ensuremath{\left|\left|{#2}\right|\right|\if\relax\detokenize{#1}\relax\else _{#1}\fi}}

\newcommand{\inverse}[1]{\ensuremath{{#1}^{-1}}}
\newcommand{\invtranspose}[1]{\ensuremath{{#1}^{\text{-T}}}}

\newcommand{\tnsrfour}[1]{\ensuremath{\mathbb{#1}}}
\newcommand{\tnsr}[1]{\ensuremath{\mathbf{#1}}}

\newcommand{\vctr}[1]{\ensuremath{\mathbf{#1}}}
\newcommand{\vctrgreek}[1]{\ensuremath{\bm{#1}}}

\newcommand{\stiffness}{\tnsrfour C}

\newcommand{\Fp}[1][]{\ensuremath{\tnsr F_\text{p}^{#1}}}

\newcommand{\Fpinv}[1][]{\ensuremath{\inverse{\Fp}}}
\newcommand{\Fpinvtranspose}[1][]{\invtranspose{\Fp}}
\newcommand{\Fe}[1][]{\ensuremath{\tnsr F_\text{e}^{#1}}}

\newcommand{\Feinv}[1][]{\ensuremath{\inverse{\Fe}}}

\newcommand{\lnameref}[1]{%
	\bgroup
	\let\nmu\MakeLowercase
	\nameref{#1}\egroup}
\newcommand{\fnameref}[1]{%
	\bgroup
	\def\nmu{\let\nmu\MakeLowercase}%
	\nameref{#1}\egroup}

\newcommand{\nmu}{}

\begin{document}
\begin{frontmatter}

\title{CALPHAD-informed phase-field modeling of grain boundary microchemistry and precipitation in Al-Zn-Mg-Cu alloys}

\author[mpie]{Chuanlai Liu\corref{cor1}}
\ead{c.liu@mpie.de}
\author[manchester]{Alistair Garner}
\author[mpie]{Huan Zhao}
\author[manchester]{Philip B. Prangnell}
\author[mpie,imperial]{Baptiste Gault}
\author[mpie]{Dierk Raabe}
\author[manchester]{Pratheek Shanthraj\corref{cor1}}
\ead{pratheek.shanthraj@manchester.ac.uk}
\cortext[cor1]{Corresponding author}
\address[mpie]{Max-Planck-Institut f\"ur Eisenforschung GmbH, Max-Planck-Str.~1, 40237 D\"usseldorf, Germany}
\address[manchester]{Department of Materials, University of Manchester, MSS Tower, Manchester M13 9PL, UK}
\address[imperial]{Department of Materials, Imperial College, South Kensington, London SW7 2AZ, UK}
\begin{abstract}
The grain boundary (GB) microchemistry and precipitation behaviour in high-strength Al-Zn-Mg-Cu alloys has an important influence on their mechanical and electrochemical properties.
\revision{Simulation} of \revision{the} GB segregation, precipitation, and solute distribution in these alloys requires an accurate description of the thermodynamics and kinetics of this multi-component system. 
\revision{CALPHAD databases have been successfully developed for equilibrium thermodynamic calculations in complex multi-component systems, and in recent years have been combined with diffusion simulations.}
In this work, we have directly  incorporated a CALPHAD database into a phase-field framework, to simulate, with high fidelity, the complex kinetics of the non-equilibrium GB microstructures that develop in these \revision{important commercial} alloys during heat treatment.
In particular, the influence of GB solute segregation, GB diffusion, precipitate number density, and far-field matrix composition, on the growth of a population of GB $\eta$-precipitates, was systematically investigated in a model Al-Zn-Mg-Cu alloy of near AA7050 composition.
It is shown that the GB solute distribution in the early stages of ageing was highly heterogeneous and strongly affected by the distribution of GB $\eta$-precipitates.
Significant Mg and Cu GB segregation \revision{was} predicted to remain during overageing, while Zn was rapidly depleted.
This non-trivial GB segregation behaviour markedly influenced the resulting precipitate morphologies, but the overall precipitate \revision{transformation} kinetics on \revision{a} GB were relatively unaffected.
Furthermore, solute depletion adjacent to the GB was largely determined by Zn and Mg diffusion, which will affect the development of precipitate free zones during the early stages of ageing.
The simulation results were compared with scanning transmission electron microscopy and atom probe tomography characterisation of alloys of the \revision{similar} composition, with \revision{good} agreement. 
\end{abstract}


\end{frontmatter}

\section{Introduction}
\label{sec: introduction}
Precipitation hardened 7xxx aluminium alloys, belonging to the Al-Zn-Mg-(Cu) system, are widely used in the aerospace sector due to their outstanding combination of high strength, low density, and high fracture toughness \citep{rao2016stress}.
However, these alloys are susceptible to localised exfoliation and intergranular corrosion in the under-aged and peak-aged temper states, which at high stress levels in certain environments can also lead to stress corrosion cracking (SCC) \citep{marlaud2011relationship,knight2010correlations}.
It is widely acknowledged that the underlying corrosion and stress corrosion mechanisms are sensitive to the local grain boundary (GB) microstructure and microchemistry \citep{GARNER2021190,kairy2018role}.
For instance, \revision{the} dissolution of GB precipitates, whereby the $\eta$-phase, \revision{which is} polarised anodically \revision{relative} to the Al matrix, \revision{can} generate micro-galvanic couples with the precipitate free zone (PFZ) \citep{knight2010correlations,kairy2018role}, has been identified as a main mechanism \revision{of} in-service stress corrosion failure in these alloys.
The substitution of Cu for Zn in the $\eta$-phase can mitigate anodic dissolution, as it reduces the eletrochemical potential between the GB precipitates and the matrix \citep{ramgopal2001electrochemical,knight2015some}.
Cu-enrichment in the $\eta$-phase is achieved in practise through alloying \citep{deschamps1999influence, marlaud2010influence}, as well as \revision{from} modified heat treatments, such as retrogression and re-ageing (RRA) treatments \citep{marlaud2010evolution}, and is accompanied by a marked reduction in intergranular stress corrosion cracking \citep{knight2010correlations, knight2015some}.

Understanding the evolution of GB microstructure is, thus, a key avenue for advancing the development and deployment of new 7xxx alloys.
However, currently the role of alloy composition and heat treatment on the GB state is still unclear \citep{rao2016stress,kairy2018role}, and a consensus on the segregation behaviour and GB precipitation kinetics is lacking, due to the confined and complex nature of GBs, particularly in 7xxx alloys which \revision{contain} multiple substitutional elements \citep{zhao2018segregation,robson2019analytical}.
This has resulted in intense efforts to characterise the GBs in 7xxx alloys at \revision{the} atomic-scale using advanced techniques, such as aberration-corrected scanning transmission electron microscopy (STEM) \citep{kairy2018role,robson2019analytical,GARNER2021190} and atom probe tomography (APT) \citep{sha2011segregation,marlaud2010evolution,zhao2018segregation,zhang2019dynamic}.
In recent work, on which we focus herein, APT \revision{revealed} significant solute \revision{segregation to} GBs, in a model as-quenched Al-Zn-Mg-Cu alloy \citep{zhao2018segregation,zhao2020interplay}, and a significant acceleration of the sequence of precipitation compared to that \revision{found} in the bulk of the grains.
Grain-size effects on segregation have also been investigated, \revision{with} higher Zn segregation \revision{being} observed in a coarse-grained alloy compared to \revision{in} an ultrafine-grained alloy, while similar Mg and Cu segregation levels were observed \citep{sha2011segregation,zhao2018segregation,zhang2019dynamic}. 
STEM investigations of a  AA7085 Al-Zn-Mg-Cu forging alloy has also revealed preferential solute segregation to GBs containing small-sized and closer-spaced precipitates compared to at GBs containing large-sized precipitates \citep{kairy2018role}.

GB precipitation and segregation has also been investigated through atomic-scale \citep{DETOR20074221,mianroodi2019atomistic}, thermodynamic \citep{xing2018solute,WAGIH2019228}, mean-field precipitation \citep{kamp2006modelling, svoboda2013formation}, and phase-field \citep{ma2003solute, heo2011phase, HEO2019262, svendsen2018finite} modelling.
In recent years, the phase-field method has emerged as a powerful tool to study microstructure evolution \citep{chen2002phase, steinbach2006multi, moelans2008introduction, wang2010phase, liu2018integrated, liu2019interaction}.
However, the majority of the applications of the phase-field method thus far have been limited to binary\revision{,} or ternary\revision{,} model alloy systems.
GB segregation in phase-field models has been formulated based on several approaches including gradient thermodynamics \cite{ma2003solute}, solute strain induced segregation \cite{heo2011phase}, and the explicit introduction of a solute-GB interaction parameter \citep{HEO2019262}, but their extension to multi-component systems relevant to engineering materials has proven challenging.
Phase-field methods for multi-component materials depend on an accurate description of the Gibbs free energy and solute \revision{mobility} activation energy, which determine the precipitate composition, volume fraction, evolution, and solute diffusion kinetics.
CALPHAD-based thermodynamic and kinetic databases have been successfully developed for this purpose, \revision{which have been widely used for equilibrium phase diagram calculations \citep{lukas2007computational} and more recently for non-equilibrium systems through their coupling with diffusion models \citep{aagren1996calculation}.}
The coupling between the phase-field model and CALPHAD databases\revision{, with solute mobility data,} has the potential to significantly advance the prediction of \revision{the evolution of complex microstructures} in engineering applications, such as the GB precipitation kinetics and segregation in the Al-Zn-Mg-Cu system.
However, direct incorporation of CALPHAD-based free energies into phase-field methods poses formidable numerical challenges\revision{,} due to their ill-posedness in the dilute limit and the complexity of the diffuse interface equilibrium condition \citep{kim1999phase} in multi-component systems. 

CALPHAD-informed phase-field modeling has been previously approached with varying degrees of fidelity.
Direct linking, through interfaces to commercial CALPHAD software, has been performed in \citep{grafe2000coupling, eiken2006multiphase}, but has proven to be extremely computationally expensive.
In \citep{zhu2004three}, the thermodynamic equilibrium conditions at the interface were pre-calculated and stored in a database, which was then queried during the phase-field simulation, while in \citep{zhang2015incorporating}, the interface equilibrium condition was replaced by a finite interface dissipation model.
While \revision{all} these approaches attempt to  accurately replicate the underlying thermodynamics, their application is limited to simple binary and ternary systems, as increasing the number of component\revision{s} rapidly escalates the computational complexity. 
In very specific cases, linking phase-field order parameters to CALPHAD site fractions can be used to reduce model complexity \citep{kitashima2009new}.
Polynomial or piecewise approximation of the Gibbs free energy are also commonly used \citep{wu2004simulating, gao2012p}, but such approaches can lead to severe errors even in binary systems \citep{hu2007thermodynamic}.

Recently, a phase-field model for multi-component systems based on a grand-potential functional was derived by Plapp \citep{plapp2011unified}.
In such an approach, the thermodynamics of the system \revision{was} reformulated in terms of the chemical potential, thus greatly simplifying the equilibrium condition at the diffuse interface \citep{kim1999phase}.
However, \revision{with this approach,} the grand-potential\revision{s} of the bulk phases \revision{were} obtained by a Legendre transformation of the Gibbs free energy, which does not exist for non-convex forms.
More recently, a time-discrete semi-analytical inversion of the thermodynamic relations, that is applicable to general forms of the Gibbs free energy, was formulated by the current authors \citep{sha2020cmame}.
In the present work, this approach \revision{was} extended to operate with a recognized CALPHAD database, enabling the investigation of GB precipitation and microchemistry evolution in \revision{a quaternary} multi-component Al-Zn-Mg-Cu system, at length and time scales relevant to typical industrial heat treatment processes.
As a first step in tackling this complex problem we have simplified the GB $\eta$-precipitate to be stoichiometric and used a fixed composition of $(\text{Zn}_{45.4}, \text{Al}_{16.8}, \text{Cu}_{4.5} )\text{Mg}_{33.3}$ taken from the APT analysis\revision{,} \revision{a}s ignoring variation in $\eta$-precipitate composition can substantially reduce the model complexity \citep{kamp2006modelling}\revision{.}
\revision{I}n this first paper this has allowed us to focus on the investigation of the effect of the GB diffusion, precipitate number density, and far-field matrix composition\revision{,} on the evolution of GB segregation and the matrix chemistry near the GB\revision{;} however, future work will include the effect of allowing the $\eta$-phase to have a range of composition.

In this paper, \cref{sec: model} begins with the description of the phase-field and thermodynamic models.
An efficient numerical implementation is then introduced, to solve the resulting formulation in terms of \revision{the} phase-field order parameters and chemical potentials.
\revision{The developed model enables the direct usage of CALPHAD thermodynamic and atomic mobility databases without any approximation and simplification, the implicit satisfaction of the thermodynamic Kim--Kim--Suzuki (KKS) \citep{kim1999phase} condition in the interface region, and addresses the numerical instability associated with the entropy of mixing contribution in the CALPHAD formalism.}
In \cref{sec: experiments}, a brief description of the experimental methods used to characterise a model Al-Zn-Mg-Cu alloy is then presented.
In \cref{sec: results}, the CALPHAD-informed phase-field model is validated and then applied to investigate GB $\eta$-precipitate growth, and the solute diffusion in this alloy.
Finally, the simulation results are discussed in \revision{comparison to data obtained from} experimental findings in \cref{sec: discussion}, and the conclusions are summarised in \cref{sec: conclusions}.

\section{Model formulation}
\label{sec: model}
In th\revision{is} section, the multi-phase-field model \revision{\citep{steinbach1996phase,tiaden1998multiphase,steinbach2006multi,eiken2006multiphase,nestler2000multi}} is briefly summarized, and the direct incorporation of CALPHAD-based Gibbs free energy forms into the phase-field model is outlined.

\subsection{The multi-phase-field model}
\label{sec: phase-field}
A multi-phase, multi-component system with $\alpha = 1 ~ ... ~ N$ phases and $i = 1 ~ ... ~ M$ components is considered.
Vector-valued phase-fields $\vctrgreek{\varphi}$ and phase compositions $\vctr{c}_\alpha$ are employed to describe the local volume fraction and solute compositions for each phase.
The order parameters and compositions fulfil the constraints,
\begin{equation}\label{eq:sum-convention}
\begin{aligned}
\sum_{\alpha}^{N} \varphi^\alpha = 1, \quad \text{and} \quad \sum_{\alpha}^{N} \varphi^\alpha c_i^\alpha = c_i,
\end{aligned}      
\end{equation}
respectively.
The Gibbs free energy functional \revision{for the} system is expressed as:
\begin{equation}\label{eq:free-energy}
\begin{aligned}
\mathcal{F}(\vctrgreek{\varphi},\nabla \vctrgreek{\varphi}, \vctr{c}_\alpha, \text{T}) = \int_{V} \bigg( f_\text{surf}(\vctrgreek{\varphi},\nabla \vctrgreek{\varphi}) + 
f_\text{\revision{bulk}}(\vctrgreek{\varphi},\vctr{c}_\alpha,\text{T})  \bigg) \text{d}V,
\end{aligned}      
\end{equation}
where V is the domain of consideration, T is the temperature, $f_\text{surf}$ and $f_\text{\revision{bulk}}$ represents the interface and \revision{bulk} free energy density, respectively.

The interface energy density is given by:
\begin{equation}\label{eq:interface-energy}
\begin{aligned}
f_\text{surf} (\vctrgreek{\varphi},\nabla \vctrgreek{\varphi}) = \sum_{\alpha \neq \beta}^{N} \frac{4 \sigma_{\alpha \beta}}{\eta_{\alpha \beta}} \bigg[ -\frac{\eta_{\alpha \beta}^2}{\pi^2} \nabla \varphi_\alpha \cdot \nabla \varphi_\beta + \varphi_\alpha \varphi_\beta \bigg],
\end{aligned}      
\end{equation}
where, $\sigma_{\alpha \beta}$ and $\eta_{\alpha \beta}$ is the interface energy and width between phase $\alpha$ and phase $\beta$, respectively.

\revision{
The bulk free energy contains both the elastic strain energy ($f_\text{\revision{elas}}$) and chemical free energy ($f_\text{\revision{chem}}$) contributions. 
Since GB $\eta$-precipitates generally do not have specific orientation relations with the matrix and mainly have incoherent precipitate-matrix interfaces \citep{unwin1969nucleation,butler1976situ,liu2010revisiting}, the elastic strain energy is only expected to play a minor role in the GB precipitation behaviour (see details in \cref{sec: elastic}).
Therefore, it is physically reasonable to neglect the mechanical driving force for the present application to the simulation of GB precipitation in Al-Zn-Mg-Cu alloys.
This simplification has also been widely employed in the mean-field GB precipitation models \citep{kamp2006modelling, svoboda2013formation}.%
}
The diffuse interface can be regarded as a mixture of the adjoining phases with equal chemical potentials, and the resulting chemical free energy density is given by: 
\begin{equation}\label{eq:chemical-energy}
\begin{aligned}
f_\text{chem} (\vctrgreek{\varphi},\vctr{c}_\alpha,\text{T}) = \sum_{\alpha}^{N} \varphi_\alpha f_\text{chem}^\alpha (\vctr{c}_\alpha,\text{T}).
\end{aligned}      
\end{equation}

The temporal and spatial evolution of the phase fields is driven by the minimization of the total free energy, $\mathcal{F}$,  through over-damped relaxation:
\begin{equation}\label{eq:Ginzburg}
\begin{aligned}
\dot{\varphi_\alpha} = - \sum_{\beta=1}^{\tilde{N}} \frac{M_{\alpha \beta}}{\tilde{N}} \bigg[ \frac{\delta \mathcal{F}}{\delta \varphi_\alpha} - \frac{\delta \mathcal{F}}{\delta \varphi_\beta} \bigg] = \sum_{\beta=1}^{\tilde{N}} \frac{M_{\alpha \beta}}{\tilde{N}}\bigg[ \sum_{\gamma=1}^{\tilde{N}} [\sigma_{\beta \gamma} I_{\beta \gamma} - \sigma_{\alpha \gamma} I_{\alpha \gamma}] + \Delta G_{\alpha \beta} \bigg],
\end{aligned}      
\end{equation}

\begin{equation}\label{eq:curvature}
\begin{aligned}
I_{\alpha \gamma} := \frac{8}{\eta_{\alpha \gamma}} \bigg[ \frac{\eta_{\alpha \gamma}^2}{\pi^2} \nabla^2 \varphi_\gamma + \varphi_\gamma  \bigg],
\end{aligned}      
\end{equation}
where \revision{$M_{\alpha \beta}$ is the interface mobility,} $\tilde{N}$ is the number of active phases, and $I_{\alpha \beta}$ are the generalised capillary terms. 

The chemical driving force, $\Delta G_{\alpha \beta}$, is calculated as the derivative of the chemical free energy, with respect to the order parameters, \ie\
\begin{equation}\label{eq:chemical-driving}
\begin{aligned}
\Delta G_{\alpha \beta} & = - \bigg[ \frac{\partial f_\text{chem}}{\partial \varphi_\alpha} - \frac{\partial f_\text{chem}}{\partial \varphi_\beta} \bigg] \\
& = f_\text{chem}^\beta(\vctr{c}_\beta) - f_\text{chem}^\alpha(\vctr{c}_\alpha) - \sum_{i=1}^{M-1} \bigg[ \mu_i \bigg(c_i^\beta(\vctrgreek{\mu}^\beta, \text{T}) - c_i^\alpha(\vctrgreek{\mu}^\alpha, \text{T}) \bigg)   \bigg], 
\end{aligned}      
\end{equation}
where \revision{the} vector-valued $\vctrgreek{\mu}^\alpha$ \revision{refers to} the chemical potentials for each phase and $\mu_i=\mu_i^{\alpha}=\mu_i^{\beta}$ is calculated relative to the solvent component $M$:
\begin{equation}\label{eq:chemical-potential}
\begin{aligned}
\mu_i^{\alpha} = \bigg( \frac{\partial f_\text{chem}^\alpha }{\partial c_i^\alpha} \bigg)_{\text{T,P}, c_j \neq c_i } - \bigg( \frac{\partial f_\text{chem}^\alpha }{\partial c_M^\alpha} \bigg)_{\text{T,P}, c_j \neq c_M }.
\end{aligned}      
\end{equation}
When $\Delta G_{\alpha \beta} = 0$, \cref{eq:chemical-driving} represents the common tangent construction, where the chemical free energy and chemical potential can be obtained from the CALPHAD database.

\subsection{Multi-component diffusion and the CALPHAD model}
\label{sec: solute-transport}
Following \citep{onsager1931,sha2020cmame}, a linear flux-force form is assumed for diffusion \revision{of each component}, 
\begin{equation}\label{eq:diffusion-equation2}
\begin{aligned}
\dot{c_i}(\vctrgreek{\mu}) = \nabla \cdot \sum_{j=1 }^{M-1} L_{ij}^{M} \nabla \mu_j.
\end{aligned}      
\end{equation}
The phenomenological chemical mobilities, $L_{ij}^M$, in \cref{eq:diffusion-equation2} are calculated as the volume-averaged chemical mobilities for each phase, $~^\alpha L_{ij}^M$,
\begin{equation}\label{eq:avg-mobility}
\begin{aligned}
L_{ij}^M = \sum_{\alpha=1}^{N} \varphi_\alpha ~^\alpha L_{ij}^M.
\end{aligned}      
\end{equation}
The phase \revision{specific} chemical mobilities, $~^\alpha L_{ij}^M$, can then be calculated from the atomic mobilities,
\begin{equation}\label{eq:phe-coef}
\begin{aligned}
~^\alpha L_{ij}^M = \sum_{k=1}^{M} \left( \delta_{jk} - c_j^\alpha \right) \left( \delta_{ki} - c_i^\alpha \right) c_k^\alpha M_k^\alpha,
\end{aligned}      
\end{equation}
where the Kronecker delta $\delta_{ij} = 1$ when $i = j$, and $\delta_{ij} = 0$ when $i \neq j$. $M_i^\alpha$ is the atomic mobility of component $i$ in the phase $\alpha$,
\begin{equation}\label{eq:atomic-mobility}
\begin{aligned}
M_i^\alpha = \Theta_i^\alpha \frac{1}{RT} \text{exp} \bigg( \frac{\Delta Q_i^\alpha}{RT} \bigg),
\end{aligned}      
\end{equation}
where $R$ is the universal gas constant; $\Theta_i^\alpha$ is the product of the atomic jump distance (squared) and the jump frequency, \revision{and} $\Delta Q_i^\alpha$ is the diffusion activation energy of component $i$.
$\Theta_i^\alpha$ is set to unity, and the activation energy, $\Delta Q_i^\alpha$, is expressed as a function of \revision{the} compositions and temperature in terms of a Redlich-Kister polynomial,
\begin{equation}\label{eq:activation-energy}
\begin{aligned}
\Delta Q_i^\alpha = \sum_{j=1}^{M} c_j Q_i^j + \sum_{p}^{M-1} \sum_{j=p+1}^{M} c_p c_j \bigg[ \sum_{r=0,1,2, ...} ~^rQ_i^{pj} (c_p - c_j)^r  \bigg],
\end{aligned}      
\end{equation}
where $Q_i^j$ is the activation energy of species\revision{,} $i$\revision{,} in pure species\revision{,} $j$, and $~^rQ_i^{pj}$ are binary interaction parameters.

Note that the composition fields, $c_i(\vctrgreek{\mu})$, in \cref{eq:diffusion-equation2} are implicit functions of chemical potentials, and the primary variables in the transport equations are the chemical potentials rather than the compositions.
However, the solution \revision{for} \cref{eq:diffusion-equation2} requires the inversion of the chemical potential relation in order to \revision{be able to} express compositions $c_i := c_i(\vctrgreek{\mu})$ for $i = 1, ..., M-1$. 
The numerical implementation of the substitutional phase and the stoichiometric phase in the chemical potential-based transport relations, and the GB segregation and diffusion will be described in the following sections.

\subsubsection{Substitutional matrix phase}
\label{sec: substitutional}
In the present work, the FCC-matrix phase is treated as a substitutional solid-solution.
The CALPHAD-based Gibbs free energy of the substitutional phase is given by,
\begin{equation}\label{eq:disorder-energy}
\begin{aligned}
\Omega f_\text{chem}^\phi = & \sum_{i=1}^M c_i ~^0G_i^\phi + RT\sum_{i=1}^M \left(c_i \text{ln} c_i\right) + ~^\text{XS}G_\text{chem}^\phi + G_\text{mag}^\phi,
\end{aligned}      
\end{equation}
where $\Omega$ is the molar volume and $~^0G_i^\phi$ is the molar Gibbs free energy for \revision{the} pure elements in a specified lattice structure.
$G_\text{mag}^\phi$ represents the magnetic contribution to the free energy, and does not need to be considered \revision{for} application to Al-Zn-Mg-Cu alloys.
The excess term, $~^\text{XS}G_\text{chem}^\phi$, expresses the non-ideal interactions between elements in the phase, and is modelled by the following Redlich-Kister polynomial:
\begin{equation}\label{eq:excess-energy}
\begin{aligned}
~^\text{XS}G_\text{chem}^\phi = & \sum_{i=1}^{M-1}\sum_{j=i+1}^M c_i c_j \left[\sum_{v=0} ~^vL_{i,j}^{\phi}(c_i - c_j)^v\right] \\
& + \sum_{i=1}^{M-2}\sum_{j=i+1}^{M-1}\sum_{k=j+1}^M c_i c_j c_k \sum_{v=1} ~^vL_{i,j,k}^\phi \left[c_v + \frac{1 - c_i - c_j - c_k}{3}\right],
\end{aligned}      
\end{equation}
where $~^vL_{i,j}^{\phi}$ and $~^vL_{i,j,k}^\phi$ denote the binary and ternary interaction parameters. They are usually expressed as linear functions of temperature.

\revision{
It is noteworthy that the up-hill diffusion arising from the miscibility gap is not expected for the current studies of GB $\eta$-precipitation in Al-Zn-Mg-Cu alloys, based on an analysis of the chemical free energy of the solution matrix.
Therefore, the energy contribution from the gradient of concentrations (\ie\ the forth order derivative with respect to concentrations in the Cahn-Hilliard model \citep{cahn1958free,sha2020cmame}) is not considered in the current work.}
Based on the chemical free energy definition in \cref{eq:disorder-energy} and \cref{eq:excess-energy}, the chemical potential for component $i$ in the disordered phase is given by:
\begin{equation}\label{eq:chem-potential-disorder}
\begin{aligned}
\mu_i^{\alpha} = ~^0G_i^\alpha - ~^0G_M^\alpha + RT \text{ln} \frac{c_i^\alpha}{c_M^\alpha} + \bigg( \frac{\partial ~^\text{XS}G_\text{chem}^\alpha}{\partial c_i^\alpha} - \frac{\partial ~^\text{XS}G_\text{chem}^\alpha}{\partial c_M^\alpha} \bigg).
\end{aligned} 
\end{equation}

Solving \cref{eq:diffusion-equation2} requires inversion of \revision{the} chemical potentials in order to express $c_i := c_i(\vctrgreek{\mu})$ for $i = 1, ..., M-1$.
This is achieved algorithmically, in a time-discrete context, through a semi-implicit splitting of the chemical potential relation:
\begin{equation}\label{eq:splitting}
\begin{aligned}
\mu_i^{\alpha} (t_n) = \check{\mu}_i^{\alpha} (t_n) + \hat{\mu}_i^{\alpha} (t_{n-1}),
\end{aligned}      
\end{equation}
into a convex contribution:
\begin{equation}\label{eq:splitting-convex}
\begin{aligned}
\check{\mu}_i^{\alpha} (t_n) = ~^0G_i^\alpha - ~^0G_M^\alpha + RT \text{ln} \frac{c_i^\alpha (t_n)}{c_M^\alpha (t_n)},
\end{aligned}      
\end{equation}
and a non-convex contribution:
\begin{equation}\label{eq:splitting-nonconvex}
\begin{aligned}
\hat{\mu}_i^{\alpha} (t_{n-1}) = \frac{\partial ~^\text{XS}G_\text{chem}^\alpha (t_{n-1})}{\partial c_i^\alpha (t_{n-1})} - \frac{\partial ~^\text{XS}G_\text{chem}^\alpha (t_{n-1})}{\partial c_M^\alpha (t_{n-1})},
\end{aligned}      
\end{equation}
where $t_n$ and $t_{n-1}$ are consecutive discrete time intervals. 
Based on this semi-implicit splitting, the compositions, $c_i^\alpha (t_n)$, can then be expressed in terms of chemical potentials by the inversion of \cref{eq:splitting-convex},
\begin{equation}\label{eq:cal-composition}
\begin{aligned}
c_m^\alpha (t_n) = \frac{\text{exp} \bigg(\frac{\mu_m^{\alpha} (t_n) - \left(~^0G_m^\alpha - ~^0G_M^\alpha \right) - \hat{\mu}_m^{\alpha} (t_{n-1}) }{RT}  \bigg) }{1 + \sum_{i=1}^{M-1} \text{exp} \bigg(\frac{\mu_i^{\alpha} (t_n) - \left(~^0G_i^\alpha - ~^0G_M^\alpha \right) - \hat{\mu}_i^{\alpha} (t_{n-1}) }{RT}  \bigg) }.
\end{aligned}       
\end{equation}

\subsubsection{The $\eta$-precipitate phase}
In the present work, the $\eta$-precipitate has been simplified as a stoichiometric compound, with little or no solubility range. 
\revision{By neglecting} variation in $\eta$-precipitate chemistry the model complexity \revision{and computation time can be substantially reduced} \citep{kamp2006modelling}.
The chemical free energy corresponding to the stoichiometric composition at a given temperature in the CALPHAD definition \revision{is} approximated \revision{in the model} by a sharp parabolic function centred at this point, 
\begin{equation}\label{eq:stoichio}
\begin{aligned}
\Omega f_\text{chem}^\text{stoi} = G_\text{const} (T) + \sum_{ i}^{M} A_i^\text{stoi} (c_i^\text{stoi} - c_{i, \text{eq}}^\text{stoi} )^2,
\end{aligned}       
\end{equation}
where $G_\text{const}$ is the chemical free energy of \revision{the $\eta$-precipitate corresponding to the stoichiometric composition obtained from} the CALPHAD database, $c_{i, \text{eq}}^\text{stoi}$ is the stoichiometric composition, $A_i^\text{stoi}$ is a penalty chosen large enough to guarantee that the free energy increases rapidly as the composition deviates from the stoichiometric composition.
The chemical potential of species $i$ of a stoichiometric phase can then be expressed as:
\begin{equation}\label{eq:stoi-potential}
\begin{aligned}
\mu_i^\text{stoi} = 2 A_i^\text{stoi}(c_i^\text{stoi} - c_{i, \text{eq}}^\text{stoi} ).
\end{aligned}       
\end{equation}
The composition can be easily obtained from the chemical potential as:
\begin{equation}\label{eq:stoi-conc}
\begin{aligned}
c_i^\text{stoi} = \frac{\mu_i^\text{stoi}}{2 A_i^\text{stoi}} + c_{i, \text{eq}}^\text{stoi}.
\end{aligned}       
\end{equation}

\subsubsection{Grain boundary segregation}
\label{sec: GBseg}
GB segregation is assumed to result from \revision{the} reduction in interface energy, $\sigma_{\alpha \beta}$ in \cref{eq:interface-energy}, resulting from solute occupation of the boundary. 
Following the  CALPHAD approach, the reduction in interface energy can be expressed by a Redlich-Kister polynomial.
In the current formulation, however, the following first order dependence is used:
\begin{equation}\label{eq:GB-seg}
\begin{aligned}
\sigma_{\alpha \beta} = \sigma_{\alpha \beta}^0 + \sum_i c_i \Delta\sigma_{\alpha \beta}^i,
\end{aligned}      
\end{equation}
where $ \sigma_{\alpha \beta}^0$ is the un-decorated GB energy, and $\Delta\sigma_{\alpha \beta}^i$ is the relaxation in GB energy due to solute-GB interaction.
Substituting \cref{eq:GB-seg} into \cref{eq:interface-energy} results in an additional segregation term to be included in the chemical potential relations, \cref{eq:chem-potential-disorder} and \cref{eq:stoi-potential}:
\begin{equation}\label{eq:seg-potential}
\begin{aligned}
\mu_i^{seg} & = \bigg( \frac{\partial f_\text{surf} }{\partial c_i} \bigg)_{\text{T,P}, c_j \neq c_i } - \bigg( \frac{\partial f_\text{surf} }{\partial c_M} \bigg)_{\text{T,P}, c_j \neq c_M } \approx \sum_{\alpha \neq \beta} \varphi_\alpha \varphi_\beta \left(\Delta\sigma_{\alpha \beta}^i - \Delta\sigma_{\alpha \beta}^M\right),
\end{aligned}      
\end{equation}
where the gradient terms in \cref{eq:interface-energy} are ignored for simplicity.

\subsection{Numerical implementation}
The inverted thermodynamic relations, \cref{eq:cal-composition,eq:stoi-conc}, are substituted into \cref{eq:diffusion-equation2} to yield the final transport equation. 
Together with \cref{eq:Ginzburg}, these form the governing equations to be solved for the phase-field order parameters, \revision{$\vctrgreek{\varphi}$}, and the component chemical potentials, \revision{$\vctrgreek{\mu}$}.
A large-scale parallel finite element solver using the PETSc numerical library \citep{Balay_etal2016} was developed to handle the discretization and numerical solution of the proposed CALPHAD-informed phase-field model, and was implemented in the freeware material simulation kit, DAMASK \citep{roters2019damask}.
While details of the solution procedure can be found in \citep{sha2020cmame}, it is worth restating here that the thermodynamic \revision{Kim--Kim--Suzuki (KKS) condition} \citep{kim1999phase} is implicitly satisfied in the current model, as the transport relations \revision{were} reformulated as a function of \revision{the} chemical potentials instead of compositions.
The stimulation setup and boundary conditions used for validation and the GB precipitation predictions are described further below in \cref{sec: validation,sec: setup}.

\section{Experimental methods}
\label{sec: experiments}
To provide realistic boundary conditions and validate the phase-field simulations, GB microstructure and microchemistry characterisation data has been used, obtained by APT and STEM analysis.

\subsection{Atom probe tomography}
\label{sec: aptexperiments}
As we are reproducing herein the results reported in \citep{zhao2018segregation}, only a summary of the experimental protocols and material are provided. 
APT analysis was preformed on a lab-cast model Al-Zn-Mg-Cu alloy, having composition 2.69 at.\% Zn, 2.87 at.\% Mg, and 0.95 at.\% Cu, with low minor impurity element levels of 0.05 at.\% Zr, 0.01 at.\% Fe and $<$ 0.01 at.\% Si, which qualifies as a high-purity variant of AA7050.
Small \SI{20}{\milli \metre} $\times$ \SI{20}{\milli \metre} $\times$ \SI{3}{\milli \metre} samples were solution treated at \SI{475}{\celsius}, followed by quenching in water and ageing at \SI{120}{\celsius}.
Ageing was interrupted at \SI{0.5}{\hour}, \SI{2}{\hour} and \SI{24}{\hour} by water quenching.
An overaged sample was also prepared by further ageing at \SI{180}{\celsius} for \SI{6}{\hour}.
Specimens were prepared from high-angle grain boundary (HAGB) regions using a dual beam FEI Helios $\text{Xe}^{+}$  plasma focused ion beam (PFIB).
APT was performed in high-voltage pulsing mode, with a pulse fraction of \SI{20}{\percent}, at a repetition rate of 250 kHz, and base temperature of  \SI{50}{\kelvin}, in a Cameca Local Electrode Atom Probe (LEAP) 5000XS instrument.
Data reconstruction and processing were carried out using the Integrated Visualization and Analysis software (IVAS).
The tomographic reconstructions were calibrated according to the crystallographic features identified on the detector hit maps.
The room temperature natural ageing time before APT analysis was approximately 2 days.

\subsection{Scanning transmission electron microscopy}
Further validation results were obtained by STEM-EDS analysis performed on a commercial AA7050-T7651 alloy \revision{of very similar composition,} \revision{containing 2.69 at.\% Zn, 2.45 at.\% Mg, and 0.94 at.\% Cu, and minor impurity element levels of 0.03 at.\% Zr, 0.03 at.\% Fe and  0.03 at.\% Si.}
\revision{The commercial AA7050-T7651 alloy has also been} subjected to a comparable heat treatment as the lab-cast alloy in the overaged state \citep{GARNER2021190}. 
Samples were prepared for TEM analysis by in-plane lift\revision{-}out from HAGB region\revision{s} using a dual beam FEI Helios $\text{Xe}^{+}$ PFIB, followed by thinning to electron transparency operating the PFIB operated at \SI{30}{\kilo\volt} with successively lower currents of \SI{1800}{\pico\ampere} and \SI{74}{\pico\ampere} and a final low energy cleaning step at \SI{5}{\kilo\volt} and \SI{24}{\pico\ampere}. 
Chemical analysis using energy dispersive X-ray spectroscopy (EDS) was performed using an FEI Talos F200 X-FEG TEM operated at \SI{200}{\kilo\volt} with a probe current of \SI{100}{\pico\ampere} and fitted with Super-X EDS detectors. 
The probe size under these conditions was measured to be approximately \SI{1}{\nano\meter}, and high resolution EDS maps were acquired with a dwell time of \SI{25}{\micro\second}. 
Prior to analysis the GB plane was tilted as vertically as possible in the foil.
Quantification of EDS spectral images was performed using the FEI Velox software package, using standard Cliff-Lorimer K-factors and without absorption correction.

\section{Results}
\label{sec: results}
\subsection{Experimental results}
\label{sec: expresults}
\begin{figure}
	\centering
	\includegraphics[width=0.7\textwidth]{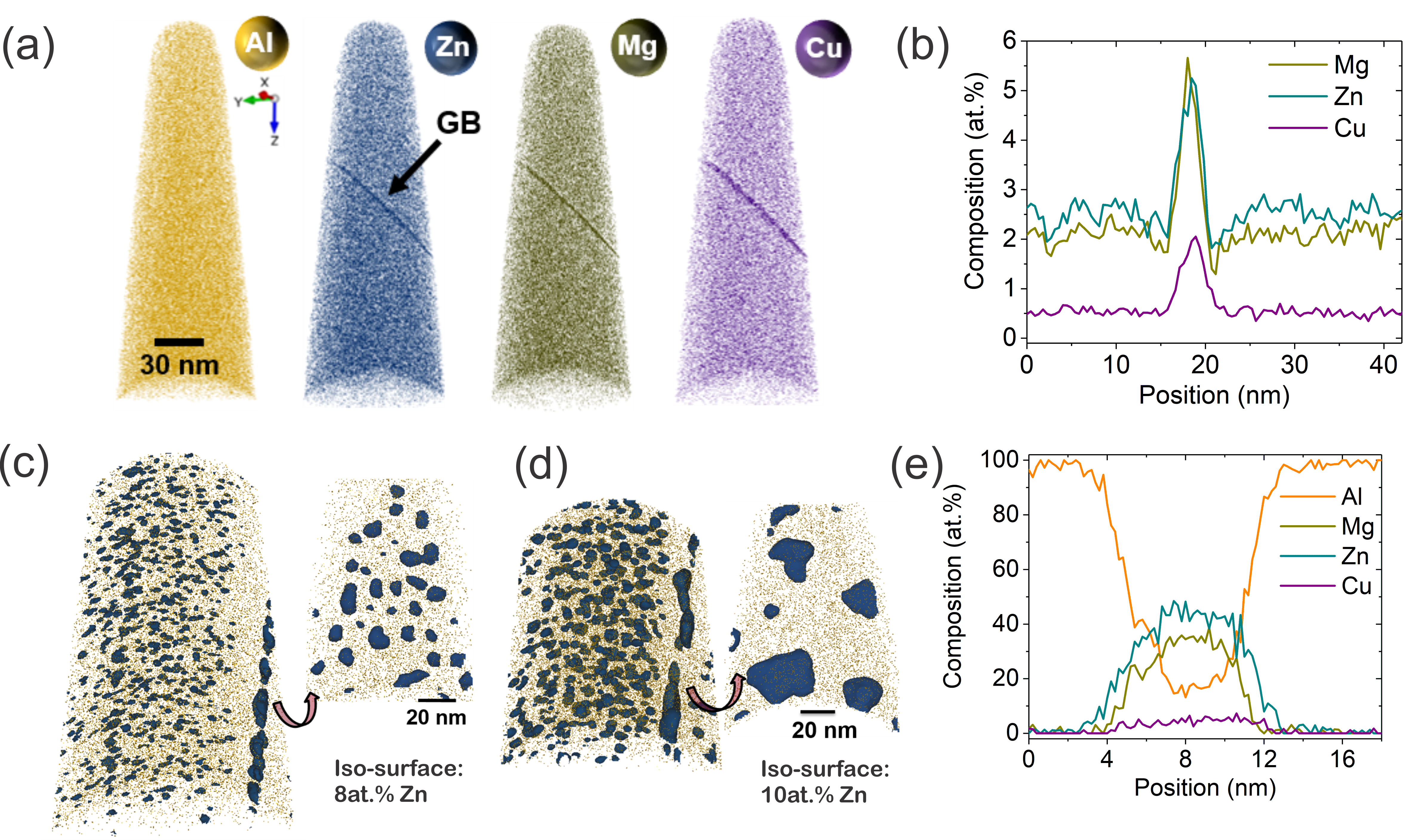}
	\caption{APT reconstruction of the GB microchemistry and precipitation in the lab-cast model AA7050 alloy \citep{zhao2018segregation}.
		(a) Atom maps of all elements in the as-quenched state, and (b) the corresponding composition profiles across the GB.
		(c) Distribution of GB precipitates after ageing for \SI{2}{\hour} at \SI{120}{\celsius}.
		(d) Distribution of GB precipitates, and (e) the corresponding compositions of the precipitate after ageing for \SI{24}{\hour} at \SI{120}{\celsius}.
		Al, Zn, Mg, and Cu are depicted in orange, dark cyan, dark yellow, and purple, respectively.
	}
	\label{fig: APT_ppts}
\end{figure}

In \cref{fig: APT_ppts}, we summarise \revision{a representative part of the } APT data initially reported in \citep{zhao2018segregation}.
\cref{fig: APT_ppts} shows the APT results \revision{for} the GB segregation and precipitation in the as-quenched Al-Zn-Mg-Cu alloy and after ageing for \SI{2}{\hour} and \SI{24}{\hour} at \SI{120}{\celsius}. 
As shown in \cref{fig: APT_ppts} (a), segregation of all the solute elements (Zn, Mg, Cu) to the HAGB was observed after quenching, and natural ageing, with relatively higher levels of enrichment of Mg and Zn (approximately 5 at.\%) and less Cu (at about 2 at.\%) found in the GB plane (\cref{fig: APT_ppts} (b)). 
The width of the enriched GB layer was measured by APT to be \SI{4}{\nano\meter}. 
\cref{fig: APT_ppts} (c, d) show examples of the evolution of the GB precipitate $\eta$-phase morphology, from spherical to more plate-shaped, after ageing for \SI{2}{\hour} and \SI{24}{\hour} at \SI{120}{\celsius} respectively, accompanied by a reduction in the precipitate number density. 
\cref{fig: APT_ppts} (e) presents the composition of the GB precipitates present in the APT samples after ageing for \SI{24}{\hour}  at \SI{120}{\celsius} (17 at.\% Al, 45 at.\% Zn, 33 at.\% Mg and 4.5 at.\% Cu).

\begin{figure}
	\centering
	\includegraphics[width=0.8\textwidth]{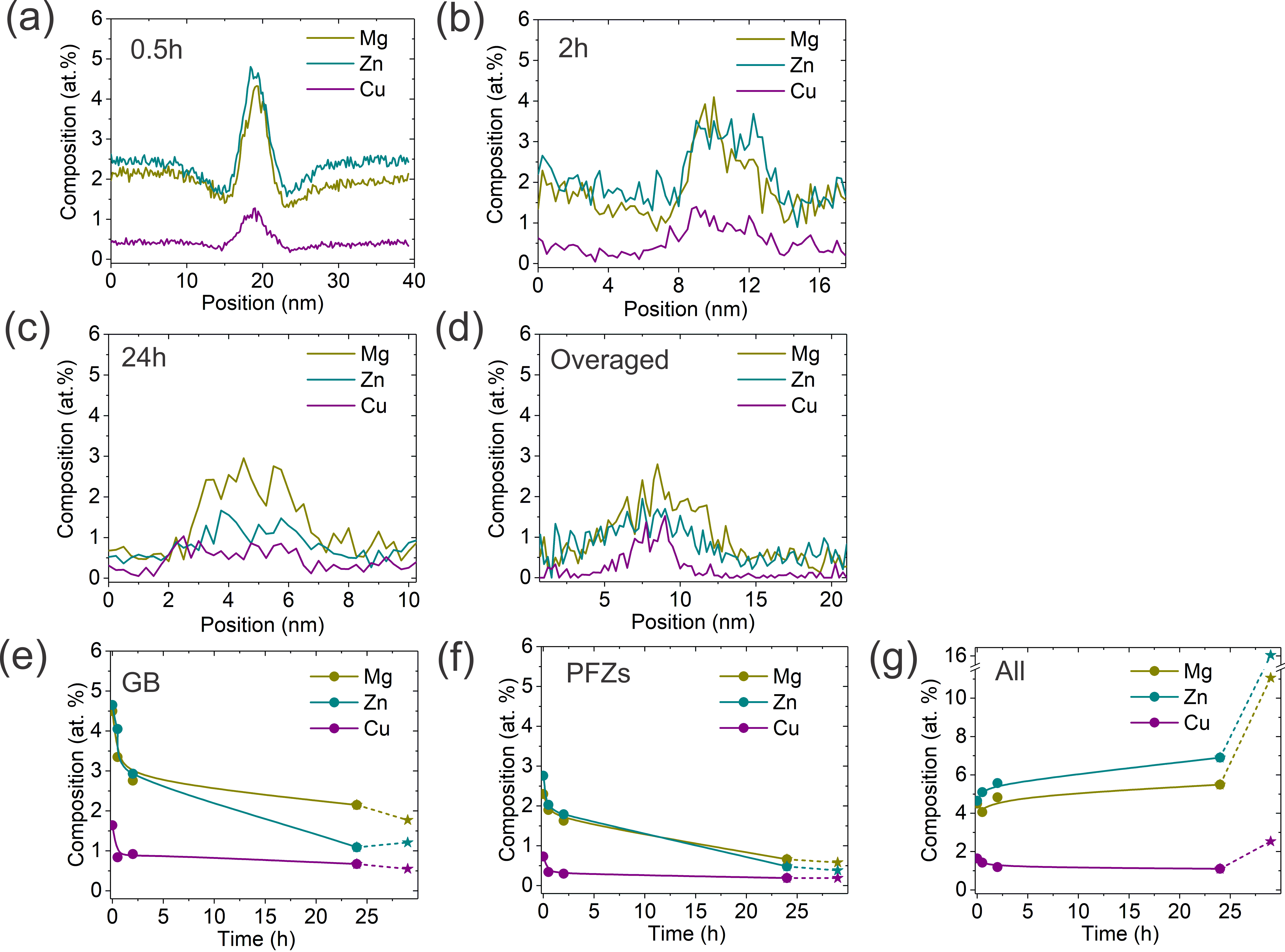}
	\caption{Summary of the evolution of the GB compositions during ageing in the lab-cast AA7050 alloy taken from the APT analysis \citep{zhao2018segregation}: 
		      example GB composition profiles after ageing for (a) \SI{0.5}{\hour}, (b) \SI{2}{\hour}, (c) \SI{24}{\hour} at \SI{120}{\celsius} and subsequently (d) \SI{6}{\hour} at \SI{180}{\celsius}, respectively.
		     (e) Average solute composition of the GB, \ie\ the average values over \SI{20}{\nano \metre} diameter cylindrical regions-of-interest across the GB without precipitates.
		     (f) Solute composition within PFZ regions.
	   	     (g) Global composition of the entire GB containing the GB precipitates.
	}
	\label{fig: APT_compositions}
\end{figure}

\cref{fig: APT_compositions} $\text{(a - d)}$ show composition profiles across the GB corresponding to different ageing conditions, that were determined as far \revision{away as possible} from (\ie\ at midway between) any observable GB precipitates.
After ageing for 0.5 h at \SI{120}{\celsius},  \cref{fig: APT_compositions} (a), segregation of Mg, Zn and Cu was observed to be similar to that in the as-quenched and naturally aged sample and a solute depletion region of width \SI{10}{\nano\meter} was observed adjacent to the GB. 
With increased ageing time at \SI{120}{\celsius} from \SI{2}{\hour} to \SI{24}{\hour}, the magnitude of the GB segregation was observed to reduce progressively and reached a minimum value after subsequent overageing for \SI{6}{\hour} at \SI{180}{\celsius}, while the region of solute depletion expanded considerably (\cref{fig: APT_compositions} (b - d)). 
Average values of the GB solute composition were measured as a function of ageing time and are shown in \cref{fig: APT_compositions} (e - g). 
\cref{fig: APT_compositions} (e) shows a clear decrease in the GB segregation levels of all elements with increased ageing time at \SI{120}{\celsius}. 
\cref{fig: APT_compositions} (f) shows the same trend in the average composition of the adjacent PFZs, with a decrease in solute composition with increased ageing time. 
Finally, the evolution of the average global GB composition, in a volume \SI{4}{\nano \metre} either side of the GB plane and including the precipitates present, is shown in \cref{fig: APT_compositions} (g). 
It can be seen that there was a gradual increase in solute composition up to the peak-aged state, due to the flux of solutes from the matrix towards the GB. 
However, the anomalously sharp increase seen in the global GB composition after overageing can be attributed to the presence of a single large precipitate in this dataset \citep{zhao2018segregation}. 

\begin{figure}
	\centering
	\includegraphics[width=0.6\textwidth]{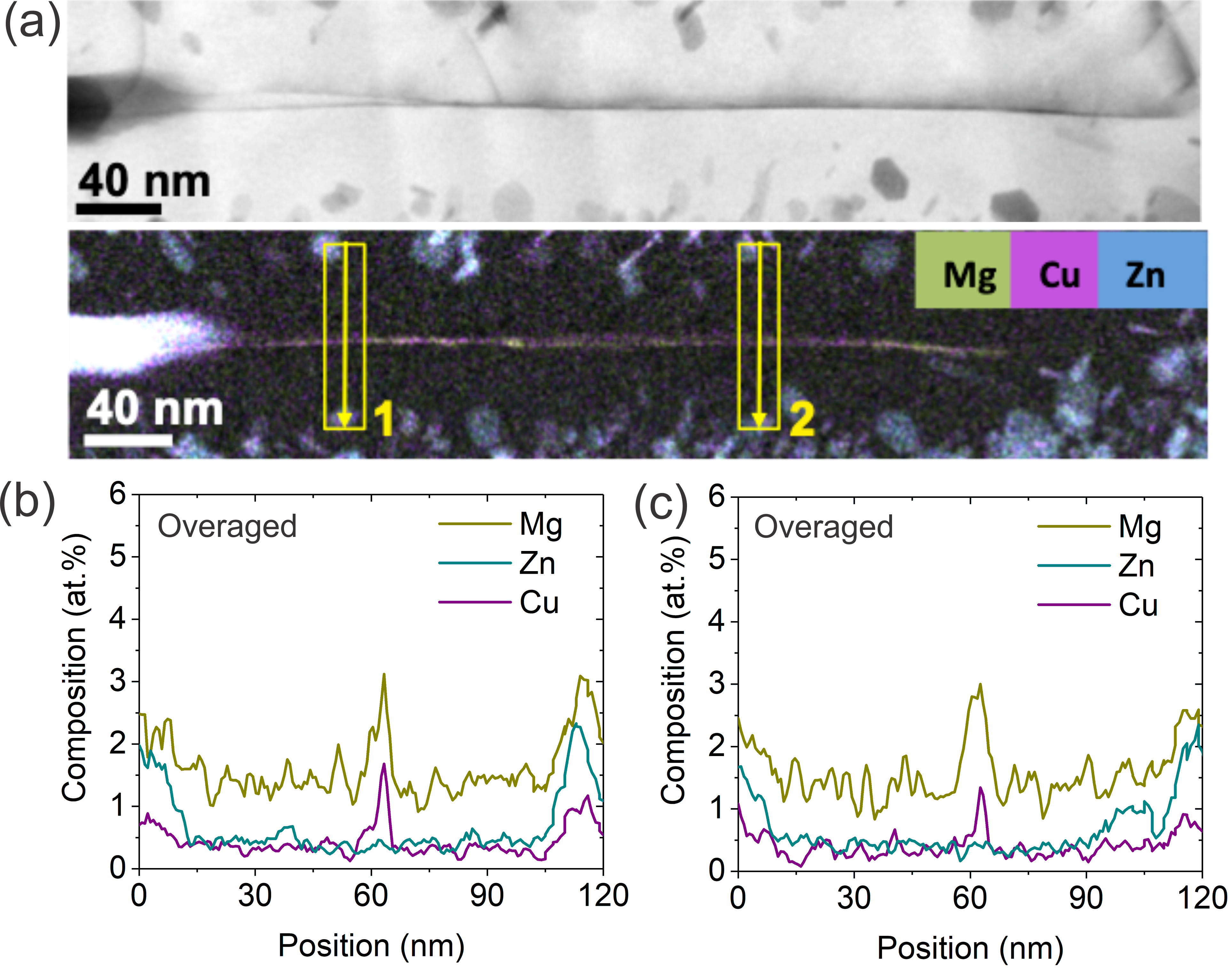}
	\caption{(a) STEM and EDS chemical maps showing typical GB segregation found in the commercial AA7050-T7651 \citep{GARNER2021190}. (b) and (c) show composition profiles across the segregation region as indicated by the position arrows in (a).
	}
	\label{fig: STEM}
\end{figure}

\cref{fig: STEM} (a) shows examples of  similar GB segregation measured in a commercial AA7050 \revision{alloy} \revision{in an} overaged \revision{T75651} temper,  \SI{140}{\milli \metre} thick plate at the T/4 position by STEM-EDS \citep{GARNER2021190}. 
A GB enriched layer, of width \SI{5}{\nano\meter}, can again clearly be seen in the STEM image next to a GB precipitate (\cref{fig: STEM} (a)), with an adjacent PFZ of \SI{80}{\nano\meter} average width. 
Significant segregation of both Mg and Cu can be observed along the GB, however little Zn segregation was observed, as shown in \cref{fig: STEM} (b, c). 
From statistical analysis of several STEM-EDS line-scans in the AA7050-T7651 commercial sample, the ratio of GB solute composition to matrix composition was observed to be highest for Cu (7.5) followed by Mg (2.6) and finally lowest for Zn (1.1). 

\subsection{Model validation}
\label{sec: validation}
The CALPHAD parameters\revision{,} describing the chemical free energy of the quaternary Al-Zn-Mg-Cu solid solution phase and the $\eta$-precipitate phase, were obtained from the \revision{open source} COST570 database \citep{saunders1998cost}, and directly used in \cref{eq:disorder-energy,eq:stoichio} without any approximation or simplification.
The kinetic parameters required to calculate the solute diffusion activation energies in \cref{eq:activation-energy} were \revision{directly} obtained from the \revision{open source} MatCalc solute mobility database \citep{kozeschnik2001matcalc}.
\revision{In the COST570 database \citep{saunders1998cost}, $\eta$-phase is described by the two sublattice model, (Al,Cu,Mg,Zn)$_2$(Al,Cu,Mg,Zn)$_1$. 
By inserting the measured composition of the $\eta$-precipitate, \textit{i.e.} $(\text{Zn}_{45.4}, \text{Al}_{16.8}, \text{Cu}_{4.5} )\text{Mg}_{33.3}$, into the sublattice model, the chemical free energy corresponding to the specific precipitate composition was obtained.}
\revision{These thermodynamic and kinetic material parameters were then directly inserted in \cref{eq:disorder-energy,eq:stoichio,eq:atomic-mobility} without any further simplification in current phase-field simulations.}
\revision{The required} thermodynamic and kinetic parameters \revision{that were used} are summarised in \cref{sec: params}.

\begin{figure}
	\centering
	\includegraphics[width=0.7\textwidth]{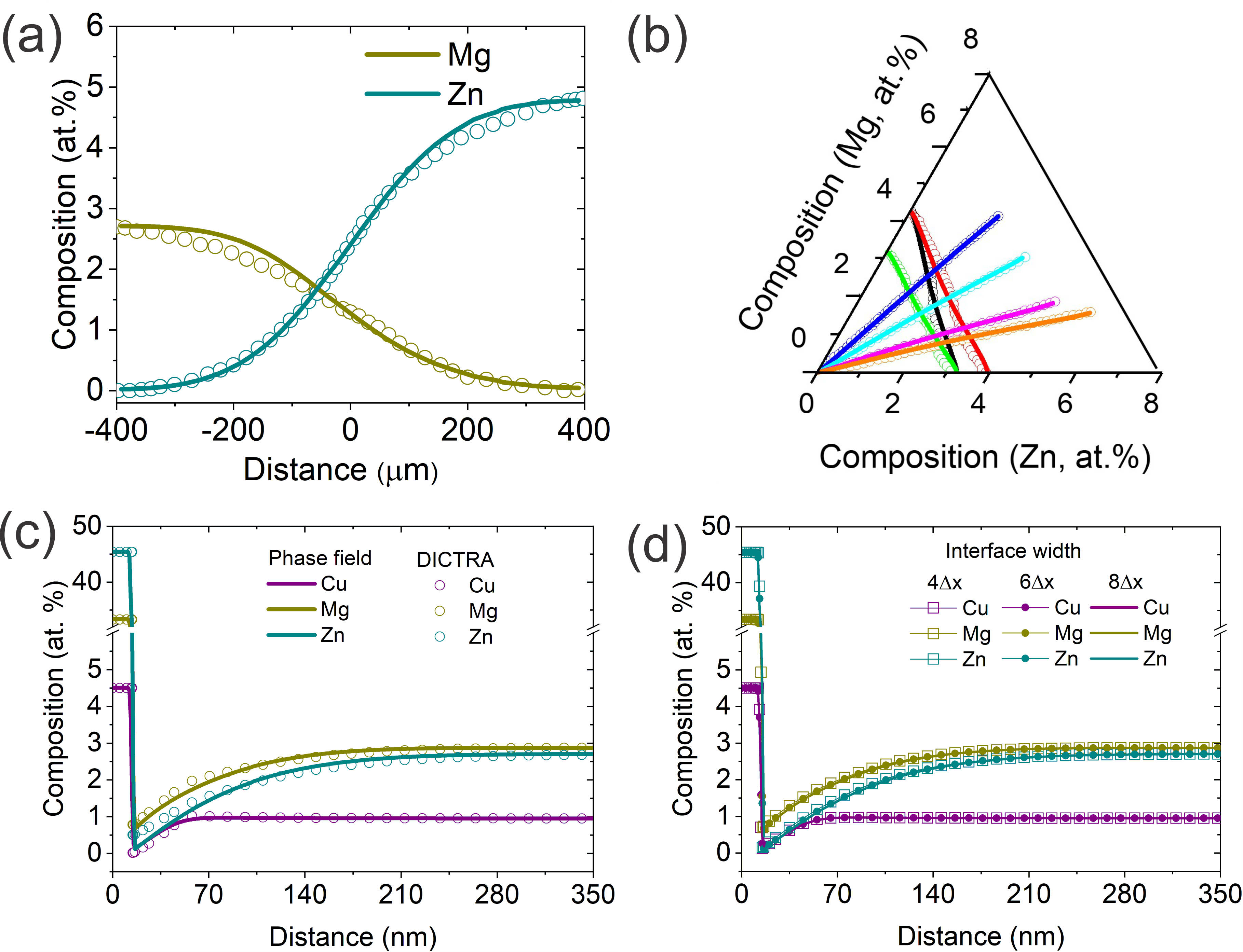}
	\caption{Preliminary validation of the phase-field model for multi-component diffusion in the Al-Zn-Mg \revision{and Al-Zn-Mg-Cu} system\revision{s}.
		(a) Solute composition profiles for a Al-4.81 at.\% Zn/Al-2.72 at.\% Mg ternary diffusion couple, after annealing for \SI{1.5}{\hour} at \SI{595}{\celsius}, obtained from phase-field simulations (solid lines) compared to experimental results (round symbols) \citep{yao2008diffusional}.
		(b) Diffusion paths for different Al-Mg-Zn ternary diffusion couples for \SI{15.8}{\hour} at \SI{482}{\celsius}. Phase-field simulation results are the solid lines and experimental results are denoted by the round symbols \citep{takahashi1999quaternary}.
		(c) Predicated \revision{Cu,} Mg and Zn diffusion profile resulting from the growth of \revision{$\eta$-precipitate ($(\text{Zn}_{45.4}, \text{Al}_{16.8}, \text{Cu}_{4.5} )\text{Mg}_{33.3}$) in an Al-2.69 at.\% Zn-2.87 at.\% Mg-0.95 at.\% Cu alloy aged at \SI{150}{\celsius} after 24h.}
		The phase-field simulations and DICTRA sharp interface simulations are represented by the solid lines and round symbols, respectively.
		\revision{(d) The corresponding phase-field simulation results with different interface widths varying from $4\Delta x$ to $8\Delta x$ ($\Delta x =$ \SI{1}{\nano \metre}). The simulation setup is the same as described in (c).} 
	} 
	\label{fig: Benchmark}
\end{figure}

The model developed was initially validated by comparing the phase-field simulations with diffusion couple experiments and precipitate growth simulations using the established DICTRA sharp interface model \citep{borgenstam2000dictra}.
\revision{In order to perform a quantitative validation, the above thermodynamic and kinetic material parameters were also used for the DICTRA sharp interface model simulations, through a user-defined database in Thermo-Calc software \citep{borgenstam2000dictra}.}
\cref{fig: Benchmark} (a) shows the composition profiles of an Al-4.81 at.\% Zn/Al-2.72 at.\% Mg ternary diffusion couple after annealing for \SI{1.5}{\hour} at \SI{595}{\celsius}.
It can be seen that the simulated results (solid lines) agree\revision{d} well with the experimental measurements (round symbols) \citep{yao2008diffusional}.
\cref{fig: Benchmark} (b) illustrates the simulated diffusion paths \revision{for} different Al-Mg-Zn ternary diffusion couples after annealing for \SI{15.8}{\hour} at \SI{482}{\celsius}, which \revision{were} in good agreement with the experimentally measured results (round symbols) \citep{takahashi1999quaternary}.
Furthermore, to validate the thermodynamic driving forces and interface kinetics of the diffuse interface model, the growth of \revision{the $\eta$-phase in a quaternary Al-2.69 at.\% Zn-2.87 at.\% Mg-0.95 at.\% Cu alloy at \SI{150}{\celsius} } was \revision{also} simulated and compared with predictions from the DICTRA sharp interface model \citep{borgenstam2000dictra}.
To perform this test comparison, the \revision{$\eta$-phase} was treated as a stoichiometric phase with a composition of \revision{$(\text{Zn}_{45.4}, \text{Al}_{16.8}, \text{Cu}_{4.5} )\text{Mg}_{33.3}$}.
A one-dimensional domain of length $L = 600 \Delta x$ was used, and a\revision{n} \revision{$\eta$-phase} of size $50 \Delta x$ was introduced at one end.
A uniform grid spacing $\Delta x = 1 \text{nm}$ and interface width $\eta = 4 \Delta x$ were used. 
As shown in \cref{fig: Benchmark} (c), \revision{after ageing for 24h at \SI{150}{\celsius}} the matrix composition profiles of \revision{Cu,} Mg and Zn accompanying the \revision{$\eta$-phase} growth simulated by the phase-field model compared favourably with the predictions of the sharp interface model in DICTRA. 
Only slight deviations in the interfacial region can be seen, attributable to the nature of the diffuse interface assumed in the phase-field model.
\revision{Moreover, the corresponding phase-field simulation results with different interface widths ($\eta = 4 \Delta x, 6 \Delta x, 8 \Delta x$, respectively) are shown in  \cref{fig: Benchmark} (d).
It can be seen that the simulated matrix composition profiles were independent of the interface width used in the phase-field simulations, which benefited from the KKS condition \citep{kim1999phase} being implicitly satisfied in the current phase-field model.
In the current study, an interface width of 4$\Delta$x instead of 8$\Delta$x was thus used, which can enable the simulation of the same total physical size with coarser finite element discretization and significantly save computation cost.}

\subsection{Simulation results}
\label{sec: GB results}
Following validation, the phase-field model was employed to investigate the evolution of the GB microstructure in a model Al-Zn-Mg-Cu supersaturated alloy, \ie\ after full solution treatment and quenching, at an isothermal temperature of \SI{120}{\celsius}.
\subsubsection{Simulation setup}
\label{sec: setup}
\begin{figure}
	\centering
	\includegraphics[width=0.6\textwidth]{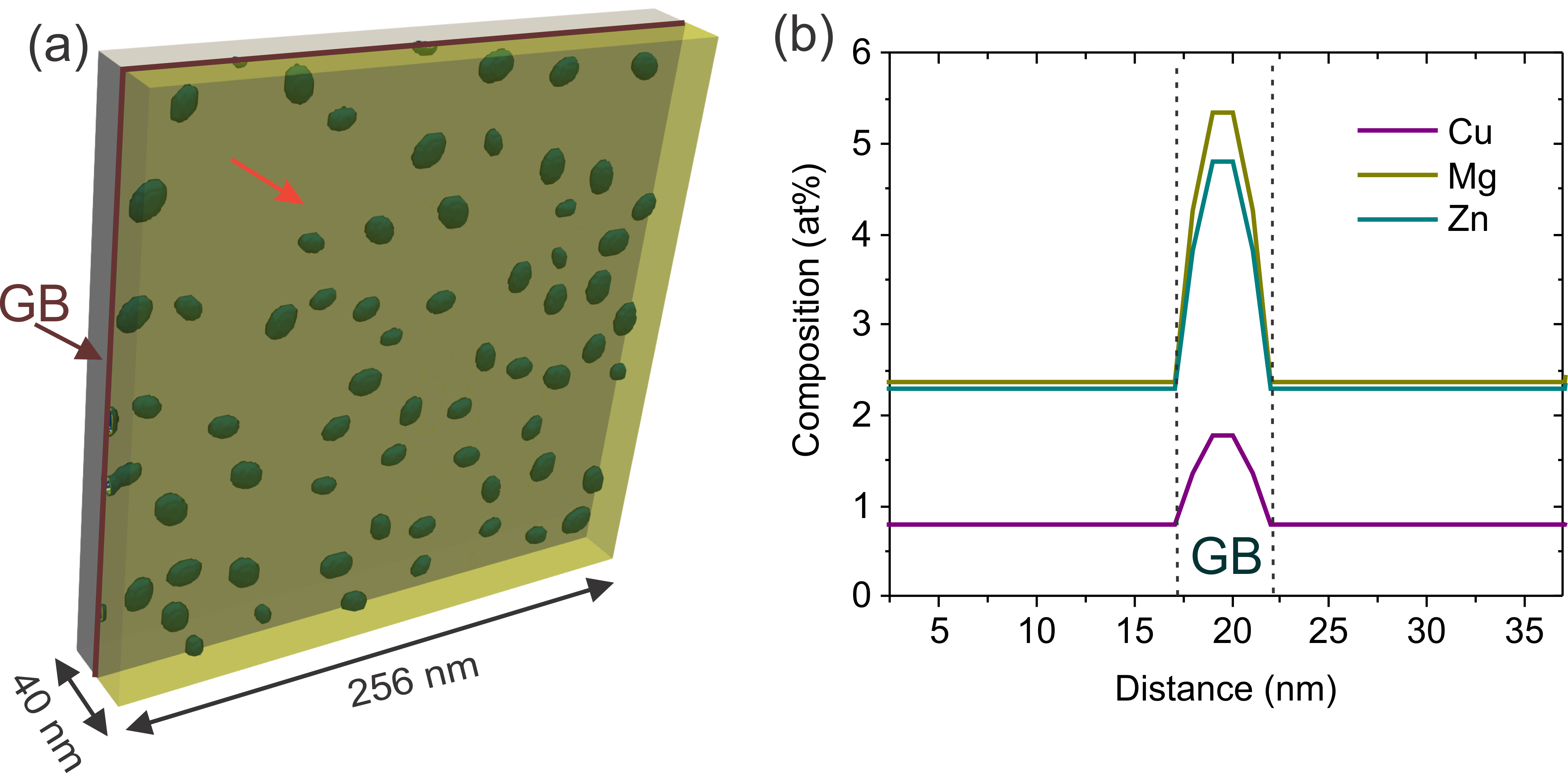}
	\caption{The APT-informed phase-field simulation model set-up.
		     (a) The bi-crystal model containing multiple precipitate nuclei on the GB, where the size and the number density information were extracted from the APT measurements.
		     The simulation box is \SI{256}{\nano \metre} $\times$ \SI{256}{\nano \metre} $\times$ \SI{40}{\nano \metre}.
		     (b) The initial solute composition profiles across the GB used in the phase-field simulations were taken from the APT results in the as-quenched condition.
	}
	\label{fig: GB_setup}
\end{figure}

The three-dimensional bi-crystal model, shown in \cref{fig: GB_setup} (a), was used for all \revision{the} simulations \revision{presented} in this section.
The domain modelled was confined in width to that of the PFZ region seen after longer ageing times and had a total width of \SI{40}{\nano \metre}, as observed in the APT experiments \citep{zhao2018segregation}.
The domain size modelled was limited to \SI{256}{\nano \metre} $\times$ \SI{256}{\nano \metre} $\times$ \SI{40}{\nano \metre}, with length scale $\Delta x = \Delta y = \Delta z = \SI{1}{\nano \metre}$ and interface width $\eta = 4\ \Delta x$. 
A time scale $\Delta t = \SI{100}{\second}$ was used.
\revision{Three phase-field order parameters were used in the present GB precipitation simulation studies, \ie\ two representing the two grains and a third representing the GB precipitate phase.}
\revision{Periodic and constant chemical potential} boundary conditions were used in the simulations.
When applying periodic boundary conditions, the average solute compositions in the simulation domain remained constant during the ageing process, representing the extreme case \revision{of} an ultrafine grain size.
In the \revision{more realistic} case \revision{of a} coarse grain size, the chemical potentials of all the solute elements (Zn, Mg, Cu) were fixed on the boundaries of the simulation box, which allow\revision{ed} solute exchange with the far-field matrix.
The GB $\eta$-precipitates were modelled as a stoichiometric phase with a composition of $(\text{Zn}_{45.4}, \text{Al}_{16.8}, \text{Cu}_{4.5} )\text{Mg}_{33.3}$ taken from the APT analysis (\cref{fig: APT_ppts} (e)).
The initial size distribution and number density of precipitates defined in the model was also informed by the APT characterisation of the early \SI{2}{\hour} aged specimen at \SI{120}{\celsius}, and a similar mean precipitate radius of \SI{5}{\nano\meter} and the number density of about \SI{1000}{\micro\meter^{-2}} was used in the simulation.
These precipitate embryos were randomly distributed in the GB plane.
A matrix composition of 2.69 at.\% Zn, 2.87 at.\% Mg and 0.95 at.\% Cu, and heat treatment temperature of \SI{120}{\celsius}, analogous to \revision{the experimental data in} \cref{sec: aptexperiments}, was simulated. 
\revision{It is noteworthy that the lab-cast alloy used for the present APT measurement represented a high-purity variant of AA7050 and experienced a much higher cooling rate during quenching due to the small specimen size. 
Therefore, quench-induced GB precipitation was not expected for the AA7050 model alloy studied, due to the much higher cooling rate compared to that normally found in a thick plate, and was thus not considered in the current phase-field simulations.}

\revision{Since GB $\eta$-precipitates in Al-Zn-Mg-Cu alloys generally do not have specific orientation relation with the matrix and the precipitate-matrix interfaces are usually incoherent \citep{unwin1969nucleation,butler1976situ,liu2010revisiting}, both precipitate-matrix interfaces and GBs were assumed to have an isotropic interface energy of \SI{0.2}{\joule\metre^{-2}} (see detail in \cref{sec: interface}).}
The interface mobility was take\revision{n} as \SI{1e-11}{\metre^{4}\joule^{-1} \second^{-1}}, which was large enough to ensure that the GB precipitate growth was a diffusion-controlled process, \revision{as} would be expected for a \revision{mainly} incoherent GB precipitate.
The molar volume of the alloy, $\Omega$, was assumed constant and had a value of \SI{1e-5}{\metre^{3}\mole^{-1}}.
The GB segregation energies in \cref{eq:GB-seg} were calibrated by fitting the \revision{initial} solute segregation profiles measured by APT across the GB in the as-quenched condition.
\cref{fig: GB_setup} (b) shows the best-fit segregation profile across the GB in the phase-field simulations, that was obtained with $\Delta\sigma_{\alpha \beta}^{\text{Zn}} =  \SI{-1800}{\joule\mole^{-1}}$, $\Delta\sigma_{\alpha \beta}^{\text{Mg}} = \SI{-2350}{\joule\mole^{-1}}$, and $\Delta\sigma_{\alpha \beta}^{\text{Cu}} = \SI{-1700}{\joule\mole^{-1}}$.
To model fast solute diffusion on the HAGB, in the simulations the solute migration energy on the GBs was decreased by \SI{20}{\percent}, consistent with measurements in similar alloy systems \citep{beke1987temperature}.
For a nominal composition of 2.69 at.\% Zn, 2.87 at.\% Mg and 0.95 at.\% Cu at \SI{120}{\celsius}, the mobility of Cu ($L_\text{CuCu}$), Mg ($L_\text{MgMg}$), and Zn ($L_\text{ZnZn}$) in the bulk, were calculated by \cref{eq:phe-coef}, to be \SI{5e-28}{\metre \squared \mole \joule^{-1} \second^{-1}}, \SI{5e-26}{\metre \squared \mole \joule^{-1} \second^{-1}}, and \SI{5.7e-26}{\metre \squared \mole \joule^{-1} \second^{-1}}, respectively, while the mobilities on GBs were \SI{1.1e-22}{\metre \squared \mole \joule^{-1} \second^{-1}}, \SI{5.5e-22}{\metre \squared \mole \joule^{-1} \second^{-1}}, and \SI{6.1e-22}{\metre \squared \mole \joule^{-1} \second^{-1}}, respectively.
This gave  an enhanced GB diffusivity by 4 to 5 orders of magnitude due to the lower migration energy.

\subsubsection{Influence of grain boundary segregation}
\label{sec: GB_segregation}
\begin{figure}
	\centering
	\includegraphics[width=0.6\textwidth]{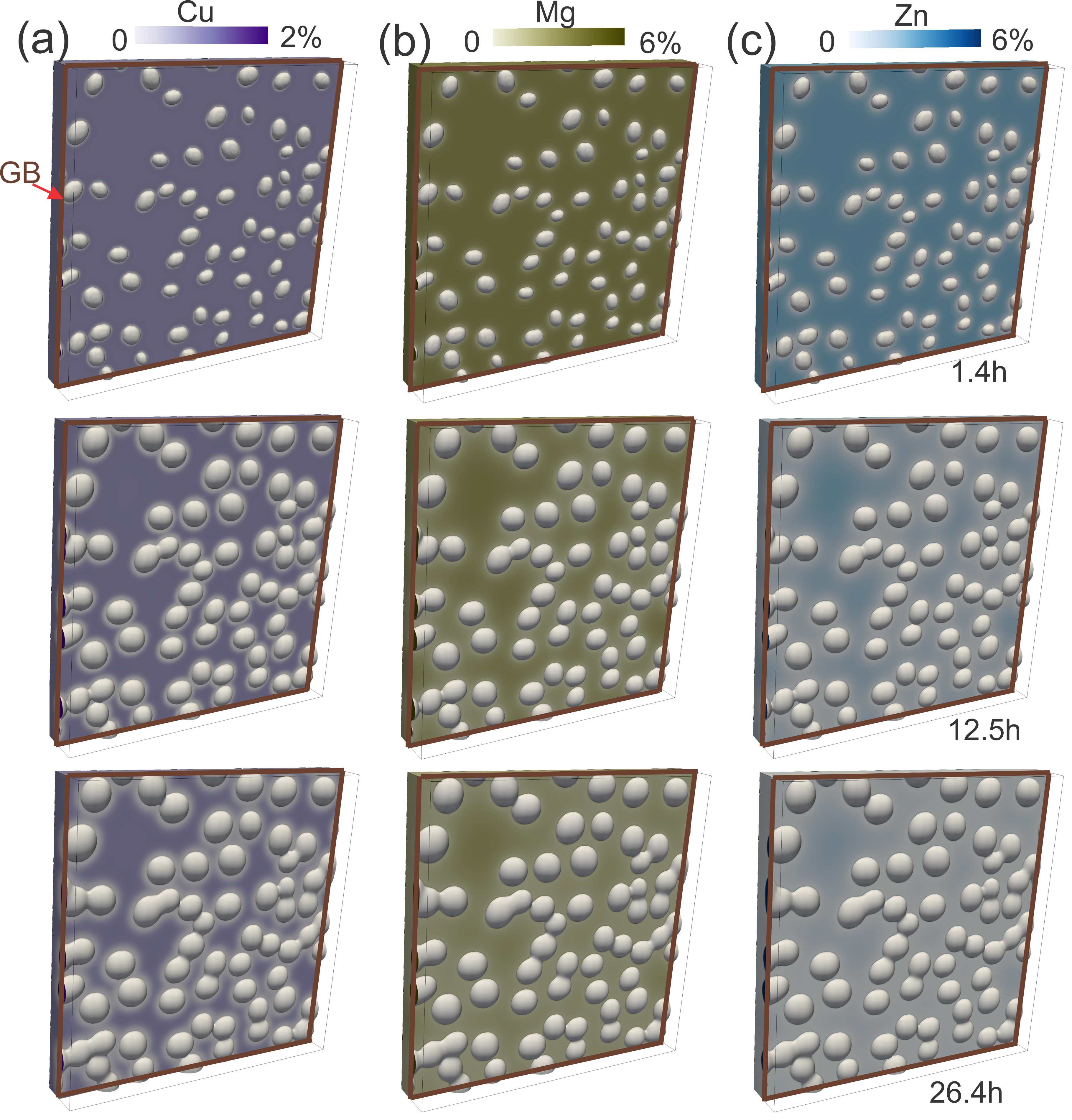}
	\caption{The simulated GB precipitate evolution when starting without GB segregation showing (a) Cu, (b) Mg and (c) Zn solute distribution on the GB plane at ageing times of \SI{1.4}{\hour}, \SI{12.5}{\hour}, and \SI{26.4}{\hour} at \SI{120}{\celsius}.
    GB diffusion was not included and assumed to be the same as the matrix.
    Periodic boundary condition was applied to the simulation box, where the average solute compositions were conserved in the simulation domain. 
	}
	\label{fig: Micro_Noseg}
\end{figure}

\begin{figure}
	\centering
	\includegraphics[width=0.6\textwidth]{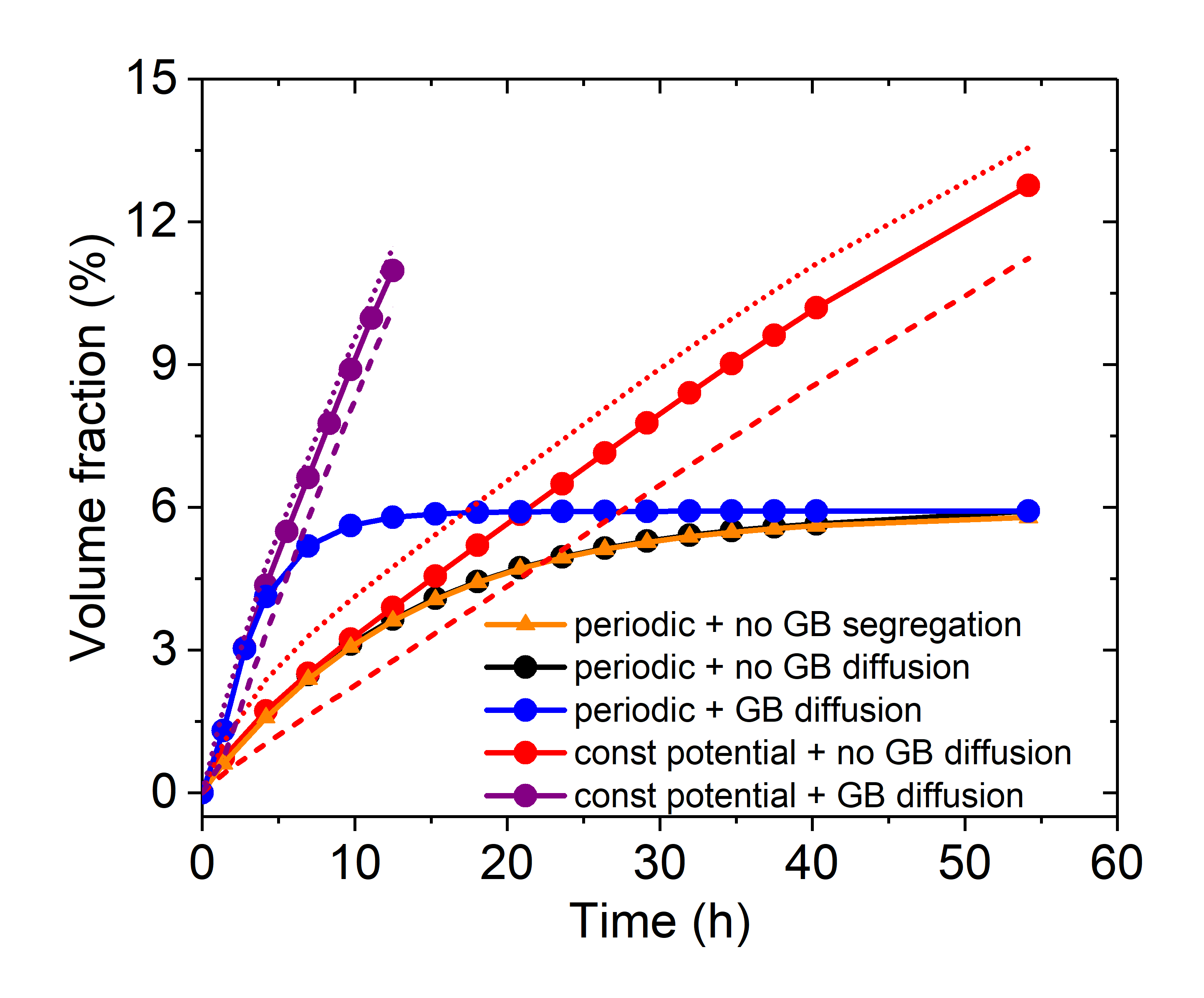}
	\caption{Influence of the GB  segregation, GB diffusion and boundary conditions on the simulated GB precipitate \revision{transformation} kinetics.
		     The dashed lines, solid lines with symbols, and dotted lines represent the results with an initial precipitate number density of \SI{500}{\micro\meter^{-2}}, \SI{950}{\micro\meter^{-2}}, and \SI{1500}{\micro\meter^{-2}}, respectively.
		     When applying periodic boundary conditions, the average solute compositions in the simulation domain remained constant during the ageing simulation.
		     Under the constant potential boundary condition, the chemical potentials of all the solute elements (Zn, Mg, Cu) were fixed on the boundaries of the simulation box, which allows solute exchange between the matrix adjacent to the GB and the far-field matrix.
		     When considering GB diffusion, solute diffusion on the GB plane was assumed to be approximately 4 to 5 orders of magnitude higher than the bulk diffusion in the simulations \citep{beke1987temperature}. 
	     }
	\label{fig: Volume_frac}
\end{figure}

Isothermal ageing simulations were first performed at \SI{120}{\celsius} without including any initial GB segregation, using the bi-crystal model described in \cref{sec: setup}.
Periodic boundary conditions were applied to the simulation domain.
For simplicity, to elucidate the initial effect of GB segregation on the GB precipitate growth kinetics, GB diffusivity was first assumed to be the same as that in the matrix.
The evolution of the GB precipitate morphology and GB plane solute composition is shown in \cref{fig: Micro_Noseg}, for \revision{simulated} ageing times corresponding to \SI{1.4}{\hour}, \SI{12.5}{\hour} and \SI{26.4}{\hour} at \SI{120}{\celsius}.
It can be seen that the depletion zone of Cu \revision{was} confined to a narrow region (approximately \SI{7}{\nano \meter} \revision{wide} at \SI{12.5}{\hour}) near precipitates, due to its relatively low diffusivity compared to Mg and Zn.
After \SI{12.5}{\hour} ageing at \SI{120}{\celsius}, the maximum compositions of Mg and Zn in the matrix \revision{were} 2.79 at.\% and 2.47 at.\% respectively, which  \revision{were lower} than the initial compositions.
This indicates that \revision{in the thin simulation box} the solute depletion zones of Mg and Zn have completely overlapped after 12.5h ageing.
With increasing ageing time, the initially rod-shaped precipitate nuclei gradually evolved into ellipsoid shapes to equilibrate the interfacial tensions between the GB and precipitate-matrix interface.
The evolution of the precipitate volume fraction with ageing times up to \SI{54.2}{\hour} at \SI{120}{\celsius} is presented in \cref{fig: Volume_frac} (orange line).
The simulated precipitate \revision{transformation} kinetics confirm that the diffusion-controlled precipitate evolution pathway follow\revision{ed} the \revision{Avrami kinetics} in the early stages, with \revision{the} growth rate \revision{rapidly reducing due to} soft impingement \revision{from} the overlap of \revision{the} solute depletion zones.

\begin{figure}
	\centering
	\includegraphics[width=0.6\textwidth]{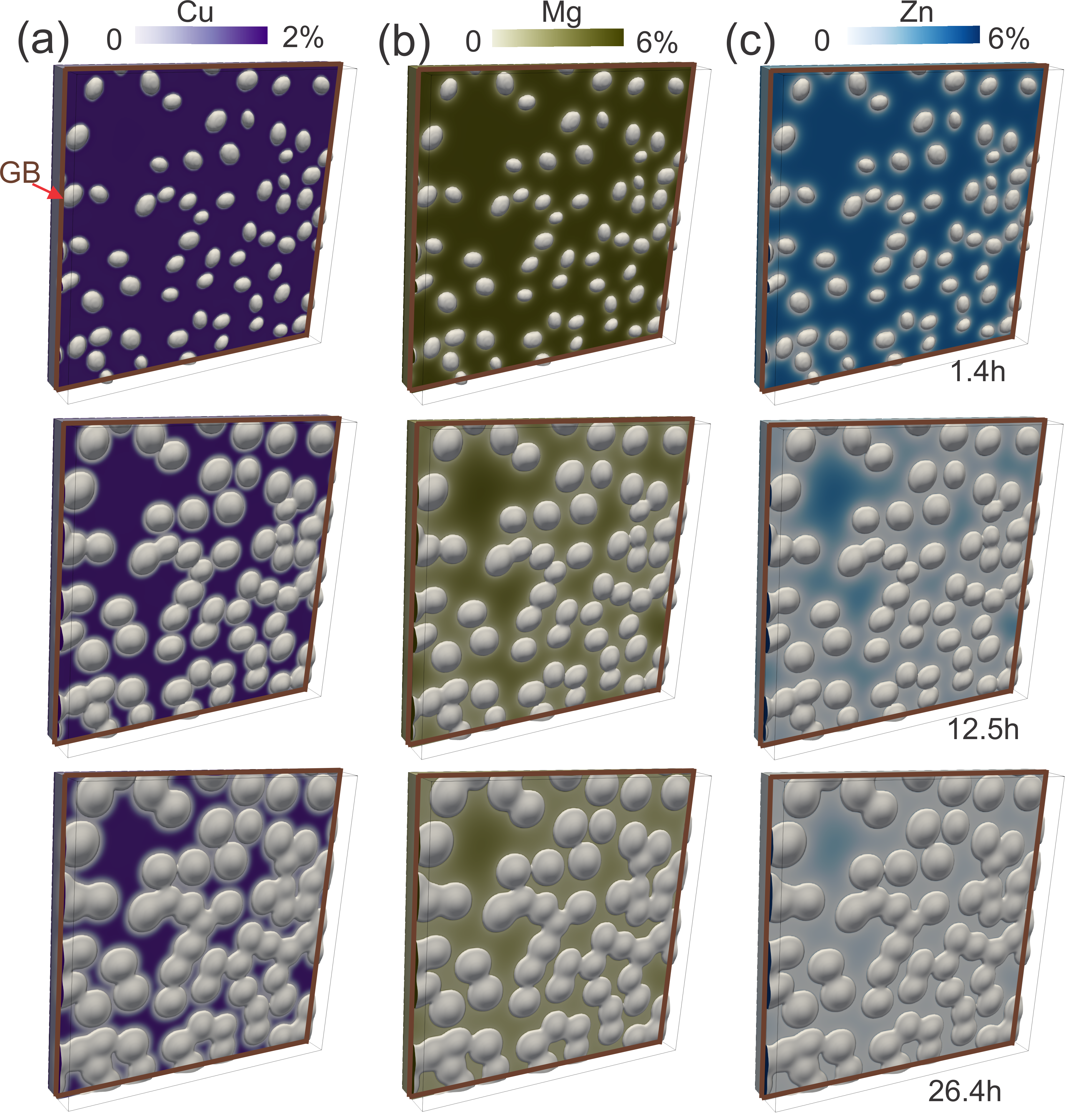}
	\caption{The effect of including GB segregation on the precipitate evolution and development of the solute field in the GB plane; (a) Cu, (b) Mg and (c) Zn solute distribution on the GB plane at ageing times of \SI{1.4}{\hour}, \SI{12.5}{\hour}, and \SI{26.4}{\hour} at \SI{120}{\celsius}.
    GB diffusion was not included and assumed to be the same as the matrix.
    Periodic boundary condition was applied to the simulation box, where the average solute compositions were conserved in the simulation domain. 
	}
	\label{fig: Micro_Seg}
\end{figure}

\cref{fig: Micro_Seg} illustrates the influence of including GB segregation on the evolution of \revision{the} GB microchemistry and precipitate morphology, where periodic boundary conditions were applied and GB diffusivity was assumed to be the same as that in the matrix.
The rod-shaped precipitate nuclei again bec\revision{a}me plate-shaped with ageing time under the influence of both solute segregation and interfacial tensions.
However, comparison of the precipitate morphologies without prior solute segregation (\cref{fig: Micro_Noseg,fig: Micro_Seg}) indicates that solute segregation result\revision{ed} in the formation of GB precipitates with a larger aspect ratio, \eg\ after ageing \SI{26.4}{\hour} at \SI{120}{\celsius} the aspect ratios \revision{were} 1.6 and 2.5, respectively.
To quantify the role of \revision{the prior} solute segregation on the precipitate growth kinetics, the evolution of the precipitate volume fraction as a function of ageing time up to \SI{54.2}{\hour} at \SI{120}{\celsius} is also depicted in \cref{fig: Volume_frac} (black line).
This show\revision{ed} that although the solute segregation affect\revision{ed} the solute partitioning and the morphological evolution of GB precipitates, it ha\revision{d} almost no influence on the average precipitate \revision{transformation} kinetics.  
This occur\revision{red} because \revision{prior} segregation to the GBs led to \revision{slightly} less solute available in the matrix adjacent to the GB, whilst the overall GB precipitate \revision{transformation} kinetics depend\revision{ed} on the average solute compositions of the simulation box.  

\begin{figure}
	\centering
	\includegraphics[width=0.8\textwidth]{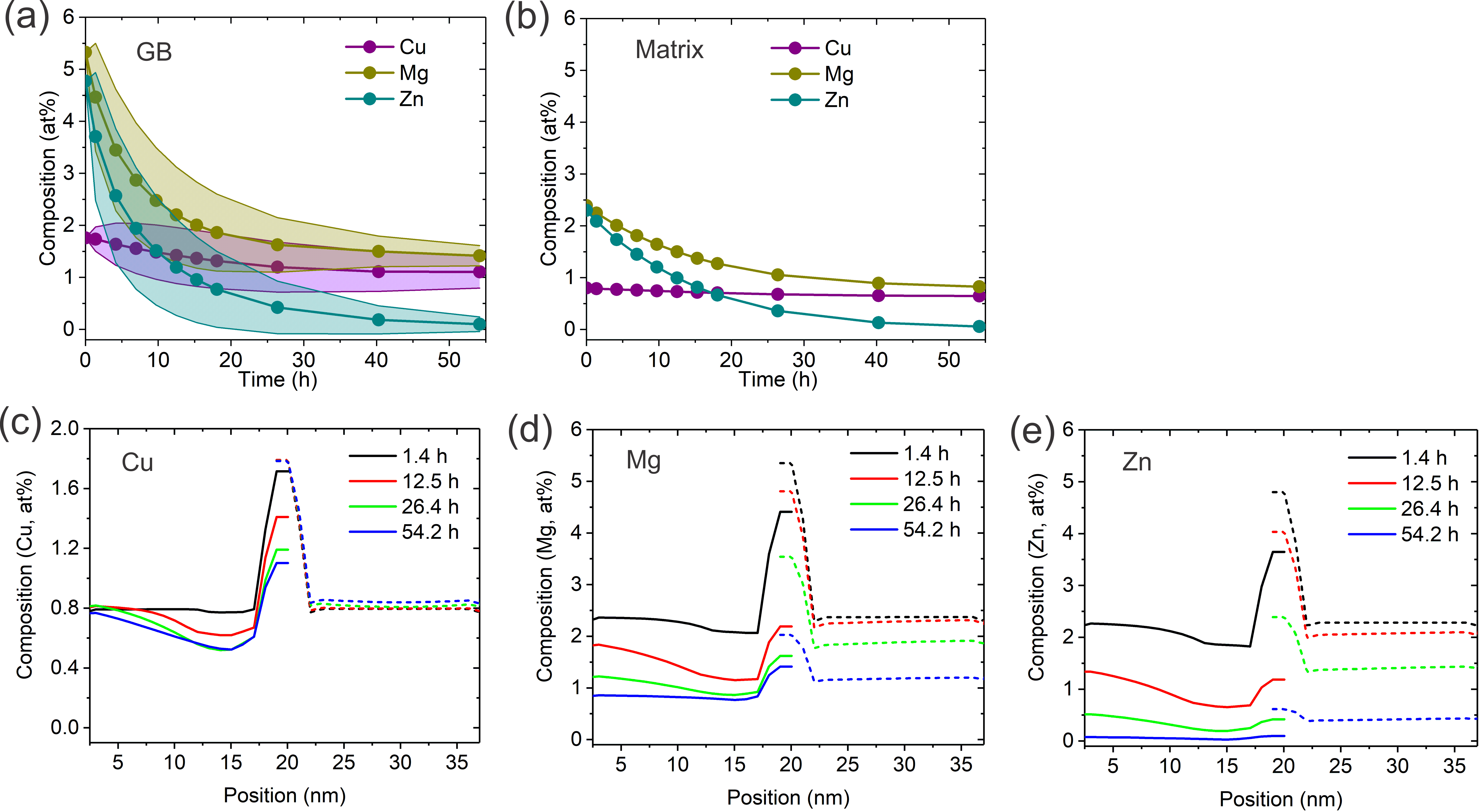}
	\caption{Evolution of the simulated average solute compositions; (a) on the GB without precipitates, with the shaded area indicating the standard deviation, and (b) within the matrix adjacent to the GB.
		Composition profiles of (c) Cu, (d) Mg, and (e) Zn across the GB with ageing time at \SI{120}{\celsius}.
		The solid lines (left half) depict the average composition across the GB and the dashed lines (right half) are the corresponding results at a local point, as far \revision{away} as possible from a precipitate, indicated by the red arrow in \cref{fig: GB_setup}.
		Only half of the profiles are plotted due to the symmetry of the results.
        GB diffusion was not included and assumed to be the same as the matrix.
		Periodic boundary condition was applied to the simulation box, where the average solute compositions were conserved in the simulation domain. 
	}
	\label{fig: Conc_profiles_Seg}
\end{figure}

Direct analysis of the predicted GB chemistry relative to the matrix levels provides more insight into the mechanical and electrochemical properties associated with the boundary regions.
The simulated evolution of the solute partitioning on the GB, as shown in \cref{fig: Micro_Seg}, reveals that Cu \revision{was expected to} strongly segregate to the GB throughout the ageing process and ha\revision{d} a relatively small depletion zone (only around \SI{10}{\nano \meter} after \SI{26.4}{\hour}). 
In contrast, the segregation magnitude of Mg and Zn decrease\revision{d} substantially with increasing ageing time and residual segregation of Zn \revision{was} not seen at the GB after \SI{26.4}{\hour} ageing treatment at \SI{120}{\celsius}. 
With this high precipitate number density, overlap of the depletion zones for both Mg and Zn can also be observed to occur a long time before that of Cu, \ie\ after ageing for only \SI{12.5}{\hour} at \SI{120}{\celsius}, due to their higher diffusivity.  
\cref{fig: Conc_profiles_Seg} (a) shows the evolution of the average compositions on the GB as a function of ageing time.
The average compositions were calculated from a small volume including the GB plane with a thickness of \SI{2}{\nano \meter}, excluding the GB plane occupied by precipitates.
This revealed that the average segregation of Mg \revision{was} predicted to decrease from 5.3 at.\% in the as-quenched state to 1.7 at.\% at \SI{26.4}{\hour} ageing, but then remained unaffected on further ageing up to \SI{54.2}{\hour} at \SI{120}{\celsius}.
In comparison, the segregation level of Zn also significantly decreased to 0.5 at.\% after \SI{26.4}{\hour} ageing, where again it then became stable, while in contrast, the segregation of Cu only exhibited a moderate decrease from 1.7 at.\% to 1.2 at.\%, over the same time-scale.
The evolution of the predicted variation in the distribution of the solute composition on the GB is also shown in \cref{fig: Conc_profiles_Seg} (a).
This reveal\revision{ed} that the solute \revision{was} predicted to distribute heterogeneously on the GB, across the whole timeframe used of \SI{54.2}{\hour} at \SI{120}{\celsius}.
The typical magnitude of the deviation for Mg and Zn was around $\pm 1$ at.\% up to \SI{12.5}{\hour} and then this gradually decreased to $\pm 0.2$ at.\% and $\pm 0.1$ at.\% at \SI{54.2}{\hour}, respectively.
However,the deviation \revision{in} the distribution of Cu was almost independent on the ageing time and had a value of around $\pm 0.4$ at.\%, since the variation in Cu was mainly confined to a narrow solute depletion zone, as shown in \cref{fig: Micro_Seg} (a).
Furthermore, the evolution of the average compositions in the whole domain not including GB precipitates was also characterised by a sharp decrease in Zn (from 2.3 at.\% to 0.06 at.\%) and Mg levels (from 2.4 at.\% to 0.8 at.\%), in contrast to only a slight reduction in the Cu level from 0.8 at.\% to 0.65 at.\%, after \SI{54.2}{\hour} ageing at \SI{120}{\celsius}, as shown in \cref{fig: Conc_profiles_Seg} (b). 

To compare the solute distributions on the GB relative to the matrix, the evolution of the simulated average composition profiles across the GB are displayed as solid lines on the left-hand side of the plot in \cref{fig: Conc_profiles_Seg} (c - e).
A clear solute depletion zone can be observed for all solute elements in the matrix near the GB.
However, Cu exhibited a considerably larger composition gradient in the matrix, even after long time ageing \revision{times} of \SI{54.2}{\hour} at \SI{120}{\celsius}, compared with Mg and Zn.
It can further be seen that the Zn profile became relatively flat after \SI{26.4}{\hour} ageing, where the segregation at the GB almost vanished.
After \SI{26.4}{\hour} ageing, \revision{modest residual segregation of} both Cu and Mg  \revision{remained} and had a similar \revision{magnitude} of 0.6\%.  
\revision{At this time} Mg \revision{was} distributed nearly homogeneously in the matrix and had a small composition gradient normal \revision{to the} GB,
whereas, owing to its lower diffusivity, depletion of Cu can be observed around the GB even after long-term ageing of \SI{54.2}{\hour} at \SI{120}{\celsius}. 
\cref{fig: Conc_profiles_Seg} (c) also shows that the width of the Cu depletion zone on either side of the GB increased from \SI{8}{\nano \meter}  to \SI{18}{\nano \meter}  when ageing from \SI{12.5}{\hour} to \SI{54.2}{\hour}.
The evolution of the width of the solute depletion zones \revision{caused by} GB precipitate growth indicates that the effect of the redistribution of Cu was restricted locally to a narrow region near the GB (\SI{18}{\nano \meter}), which \revision{from the APT data} would be within the width of the PFZ expected at \revision{an} ageing temperature of \SI{120}{\celsius}.

The dashed lines on the right side of the graphs in \cref{fig: Conc_profiles_Seg} (c - e) describes the evolution of the solute composition profiles across the GB in the simulation box at the point indicated by the red arrow in \cref{fig: GB_setup}, which is as far \revision{away} as \revision{possible} from any seed particles.
It can be seen that at this location the local distribution (dashed lines) of all the solutes in both the matrix adjacent to the GB and \revision{in the} GB exhibited different characteristics compared to the average distributions (solid lines), which illustrates the \revision{significant} heterogeneity \revision{in} the solute distribution found in both the matrix adjacent to the GB and GB plane.
For both Mg and Zn, the local composition profiles in the matrix were relatively flat and had much higher values compared to the average compositions during the whole ageing process.
Moderate local residual segregation of both Mg (0.8 at.\% excess) and Zn (0.2 at.\% excess) can also be observed after \SI{54.2}{\hour} ageing at \SI{120}{\celsius}, which was higher than the average behaviour along the GB.
In comparison, the magnitude of the local GB Cu segregation hardly changed at all with increasing ageing time. 
At this local position \revision{furthermost from any particles}, no obvious solute depletion of Cu can be observed even after long-term ageing \revision{for} \SI{54.2}{\hour} at \SI{120}{\celsius}.

\subsubsection{Influence of grain boundary diffusion}
\label{sec: GB_diffusion}

In the previous simulations, solute diffusion on GBs was assumed to be the same as that in the matrix, which represent\revision{ed} \revision{a} simplified case  as it is known that GB diffusion can significantly affect the growth behaviour of GB precipitates by, for example, the collector plate mechanism \citep{kamp2006modelling}.
Therefore, to model fast GB diffusion, the solute migration energy on GB was decreased by \SI{20}{\percent} in the current simulations, to be consistent with measurements on similar alloy systems \citep{beke1987temperature}, which report\revision{ed} an enhanced GB diffusivity by 4 to 5 orders of magnitude, as described in \cref{sec: setup}.

\begin{figure}
	\centering
	\includegraphics[width=0.6\textwidth]{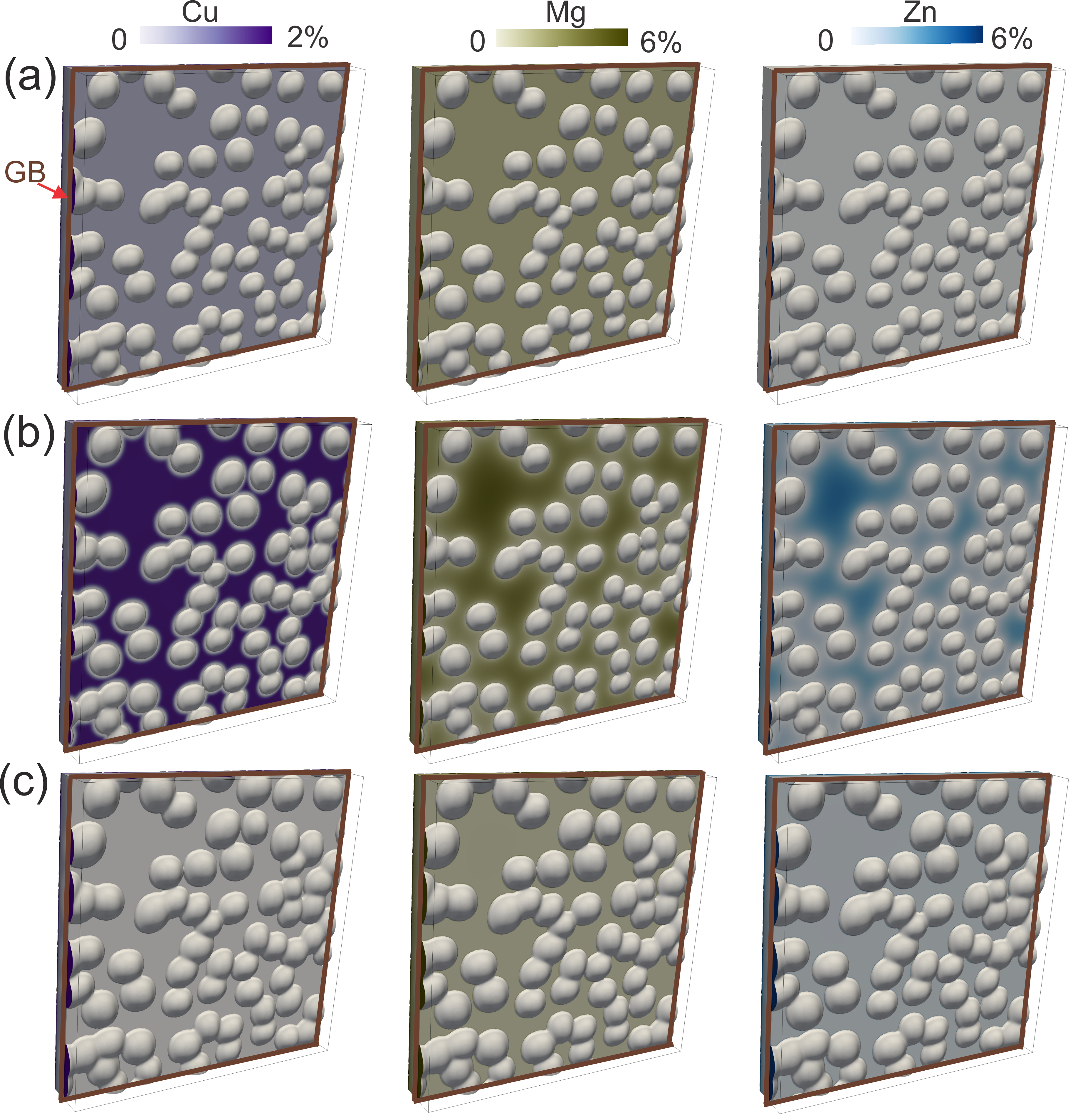}
	\caption{GB precipitate and the solute distribution at an ageing time of \SI{12.5}{\hour} at \SI{120}{\celsius}.
		(a) Periodic boundary condition was applied, where the average solute compositions were conserved in the simulation domain. 
		      GB diffusion was assumed to be approximately 4 to 5 orders of magnitude higher than the bulk diffusion.
		(b) Constant potential boundary condition was applied, where the chemical potentials of all the solute elements (Zn, Mg, Cu) were fixed on the boundaries of the simulation box, which allows the solute exchange between the matrix adjacent to the GB and the far-field matrix.
		GB diffusion was not included.
		(c) Constant potential boundary condition was applied and GB diffusion was assumed to be approximately 4 to 5 orders of magnitude higher than the bulk diffusion.
	}
	\label{fig: Micro_Seg_combine}
\end{figure}

\cref{fig: Micro_Seg_combine} (a) shows the morphology of the GB precipitates and the solute distribution on the GB plane after \SI{12.5}{\hour} ageing at \SI{120}{\celsius}, where both fast GB diffusion and solute segregation \revision{were} considered.
In these simulations, periodic boundary conditions were again applied.
It can be now noted that with the same initial distribution of precipitate seeds, the solute elements \revision{were} distributed more homogeneously on the GB \revision{plane} and had a significantly lower \revision{magnitude of} GB segregation, compared to the results when GB diffusion was not included (shown in \cref{fig: Micro_Seg}).
The effect of the GB diffusion on the predicted precipitate \revision{transformation} kinetics is shown in \cref{fig: Volume_frac} (blue line), where it can be seen that including GB diffusion significantly accelerated the precipitate \revision{transformation rate}.
The precipitate volume fraction increased more rapidly with ageing time and saturated at around \SI{11}{\hour} ageing, compared to more than \SI{40}{\hour} if GB diffusion was not considered.
After \SI{12.5}{\hour} ageing, the GB precipitate coverage also increased from \SI{33.7}{\percent}\revision{,} in the case without GB diffusion\revision{,} to \SI{41.6}{\percent} with GB diffusion, and the thickness of the GB precipitates increased from \SI{6}{\nano \meter} to \SI{9}{\nano \meter}.
This indicates that fast diffusion along the GB can substantially accelerate precipitate growth along\revision{,} as well as normal to\revision{,} the GB.

\begin{figure}
	\centering
	\includegraphics[width=0.8\textwidth]{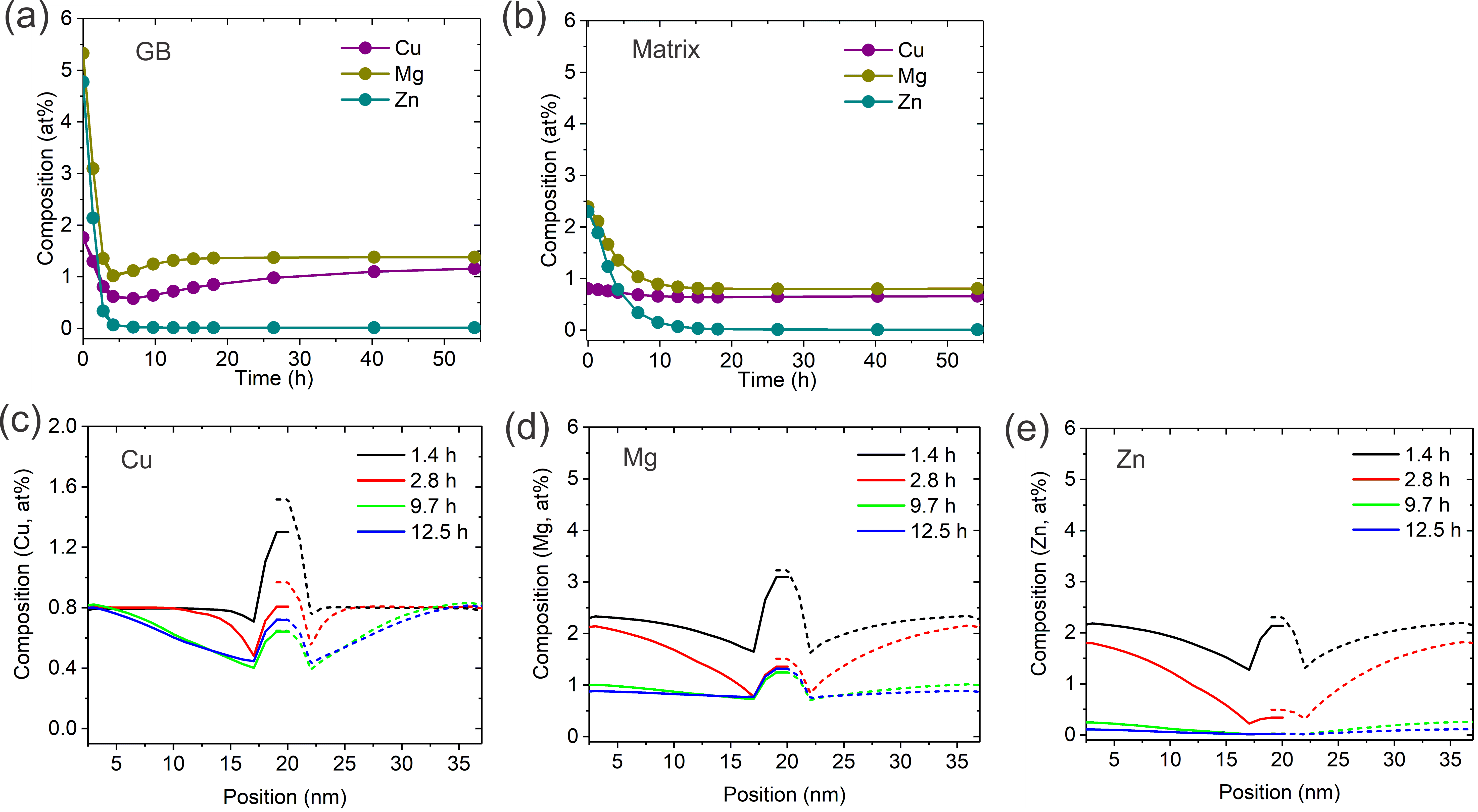}
	\caption{The effect of GB diffusion on the evolution of average solute composition (a) on the GB and (b) within the matrix adjacent to the GB.
		Composition profiles of (c) Cu, (d) Mg, and (e) Zn across the GB with ageing time at \SI{120}{\celsius}.
		The solid lines (left half) depict the average composition across the GB and the dashed lines (right half) are the corresponding results at a local point, as far \revision{away} as possible from a precipitate, indicated by the red arrow in \cref{fig: GB_setup}.
		Only half of the profiles are plotted due to the symmetry of the results.
		GB diffusion was included and assumed to be 4 to 5 orders of magnitude higher than the bulk diffusion.
		Periodic boundary condition was applied to the simulation box, where the average solute compositions were conserved in the simulation domain. 
	}
	\label{fig: Conc_profiles_Seg_GBM}
\end{figure}

Comparison of the evolution of GB chemistry (\cref{fig: Conc_profiles_Seg} (a, b) and \cref{fig: Conc_profiles_Seg_GBM} (a, b)) further demonstrates that the GB diffusion significantly enhance\revision{d} the rate of solute redistribution between the GB and matrix region, while, as expected, it ha\revision{d} negligible effect on the steady-state results after long-term \SI{54.2}{\hour} ageing at \SI{120}{\celsius}.
In both cases, as described previously, moderate GB segregation of Mg and Cu (approximately 0.6 at.\% excess) can be observed, while Zn was completely depleted in both the matrix region and GB plane.
However, comparison of \cref{fig: Conc_profiles_Seg,} (a) and \cref{fig: Conc_profiles_Seg_GBM} (a) shows that under the influence of GB diffusion, the segregation magnitudes of all the solutes exhibited a much sharper decrease after shorter ageing times (\SI{4.2}{\hour} at \SI{120}{\celsius}).
Zn also became completely depleted in  the GB earlier, while the segregation of Mg and Cu then gradually increased further with ageing time before approaching a steady lower limit taking around  \SI{11}{\hour} and \SI{40}{\hour}, respectively.
Furthermore, GB diffusion resulted in much smaller variation \revision{in the solute concentration of} all the elements on the GB (\eg\ 0.1 at. \% for Cu), as shown in \cref{fig: Conc_profiles_Seg_GBM} (a).

In \cref{fig: Conc_profiles_Seg_GBM} (c - e) the evolution of average (solid line) and local composition profiles across the GB (dashed line) \revision{were} again compared with increasing ageing time. 
The width of the Cu depletion zone was still restricted close to the GB and within the model domain of  \SI{40}{\nano \metre}, despite the introduction of higher GB diffusivity.
However, the Mg and Zn depletion zones now extended beyond the domain width after only ageing \revision{for \SI{2.8}{\hour}}.
Although \revision{some} residual GB segregation of Mg and Cu can be observed throughout the ageing treatment, the Zn GB excess disappeared more rapidly \revision{and was absent} after only \SI{2.8}{\hour}.
From the evolution of the composition profiles \revision{shown} across the GB at a local point f\revision{u}rthest from the precipitates (the red arrow in \cref{fig: GB_setup}), presented in \cref{fig: Conc_profiles_Seg_GBM} (c - e) by the dashed lines, 
it can be seen that, in comparison to the average composition profiles, they  now rapidly converged and only showed a slight deviation at the start of the ageing time.
This implies that the solutes in the matrix adjacent to the GB can be effectively \revision{``}collected\revision{'' } when they reach\revision{ed} the GBs by rapid short-circuit diffusion to the GB precipitates.
\revision{With a reduction in the migration energy for GB diffusion by 20\%,} \revision{t}here was also significantly increased parity between the GB solute mobilities of different species compared to those in the matrix, with $L_\text{ZnZn}^{GB}/L_\text{CuCu}^{GB}\approx 5$ compared to $L_\text{ZnZn}^{FCC}/L_\text{CuCu}^{FCC}\approx 100$, mitigating the diffusional bottleneck resulting from the slow diffusion of Cu.

\subsubsection{Influence of the far-field matrix}
\label{conc_flux}
The above results show that when using such a thin slab-shaped model box solute diffusion accompanying precipitation interact\revision{ed} with the domain boundary early on in the ageing process, resulting in soft impingement and th\revision{is} effectively limit\revision{ed} the growth kinetics of the GB precipitates at longer ageing times.
In this section, we therefore explore the impact of the far-field matrix on the evolution of the GB microchemistry and precipitates by allowing solute exchange between the far-field matrix and the region adjacent to the GB, through adopting \revision{a} boundary condition \revision{where} the chemical potentials on the boundary of the simulation box \revision{were fixed}.
In reality, \revision{during artificial ageing} precipitation would also occur in the matrix of a grain in competition with GB precipitation, which \revision{was} not included in the current simulation, and the real behaviour would \revision{therefore} fall between these two extreme boundary conditions. 

\cref{fig: Micro_Seg_combine} (c) shows the effect of allowing far-field solute exchange \revision{with} the GB region on the morphology of the GB precipitates and solute distribution on the GB after \SI{12.5}{\hour} ageing at \SI{120}{\celsius}, with GB diffusion.
Comparison with the periodic boundary condition at \SI{12.5}{\hour} (\cref{fig: Micro_Seg_combine} (a) and (c)) indicates that the solute flux from the far-field matrix to the GB region \revision{has} enhanced the growth \revision{rate} of \revision{the} precipitates, both along and across the GB. 
As a results, the GB precipitate coverage increased from \SI{41.6}{\percent} to \SI{55.4}{\percent} and the thickness of the precipitates increased from \SI{9}{\nano \meter} to \SI{12}{\nano \meter}.
As depicted in \cref{fig: Volume_frac}, the far-field matrix also had a significant influence on the average precipitate \revision{transformation} kinetics after \SI{6}{\hour} ageing, when the domain boundary started to interact with the range of solute depletion which inhibit\revision{ed} the GB precipitate growth (comparing the blue and purple lines), and this effect \revision{was further} compounded by faster GB diffusion (comparing the red and black lines).

\begin{figure}
	\centering
	\includegraphics[width=0.8\textwidth]{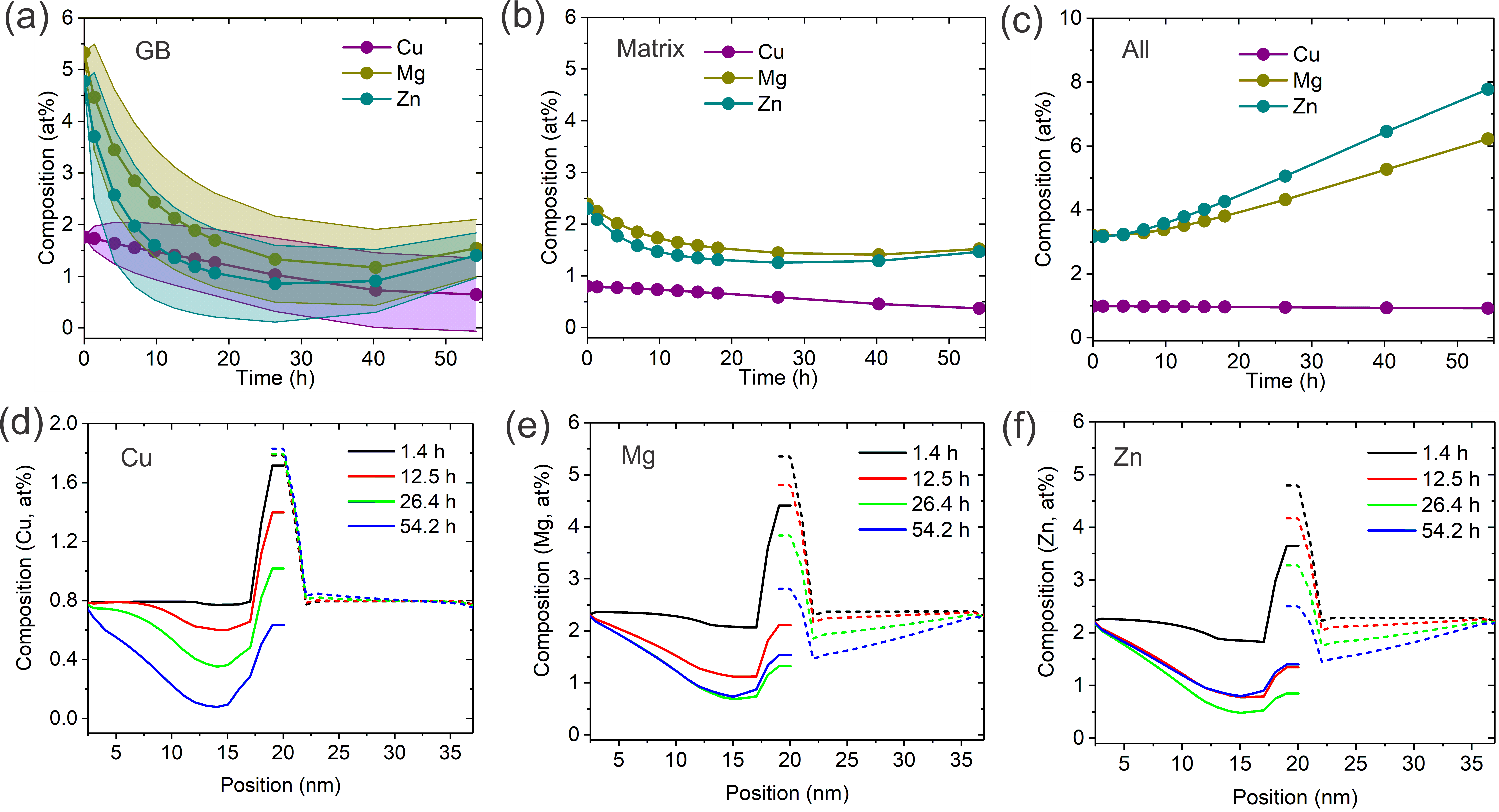}
	\caption{The effect of constant potential  boundary conditions on the evolution of the average solute composition; (a) on the GB, with the shaded area indicating the standard deviation, (b) within the matrix adjacent to the GB, and (c) in the overall domain.
		Composition profiles of (d) Cu, (e) Mg, and (f) Zn across the GB with ageing time at \SI{120}{\celsius}.
		The solid lines (left half) depict the average composition across the GB and the dashed lines (right half) are the corresponding results at a local point, as far \revision{away} as possible from a precipitate, indicated by the red arrow in \cref{fig: GB_setup}.
		Only half of the profiles are plotted due to the symmetry of the results.
		GB diffusion was not included and assumed to be the same as the matrix.
		The chemical potentials of all the solute elements (Zn, Mg, Cu) were fixed on the boundaries of the simulation box, which allows the solute exchange between the matrix adjacent to the GB and the far-field matrix.
		}
	\label{fig: Conc_profiles_fix_seg}
\end{figure}

The influence of a far-field matrix solute flux on the evolution of the solute compositions on the GB plane and within the matrix close to the GB is shown in \cref{fig: Conc_profiles_fix_seg} (a - c), first without GB diffusion.
As shown in \cref{fig: Conc_profiles_fix_seg} (a), the GB segregation of Mg and Zn exhibited a sharp decrease at the beginning of ageing, as seen previously, but this no longer bottomed out across the simulation box and was followed by a gradual increase \revision{for longer times}, while the GB Cu segregation still decreased almost  linearly  with ageing \revision{time}, as it \revision{was} less influenced by the model boundary conditions due to its lower diffusivity.
The evolution of the solute content in the matrix close to the GB (\cref{fig: Conc_profiles_fix_seg} (b)) \revision{was} also characterised by an initial rapid decrease of Mg and Zn followed by a slight increase after long\revision{er} time ageing.
A constant decrease of the average Cu composition in the matrix adjacent to the GB \revision{was} observed. 
However, \cref{fig: Conc_profiles_fix_seg} (c) now shows that the global composition of Mg and Zn in the overall domain increased slowly when ageing from \SI{12.5}{\hour} to \SI{54.2}{\hour} at \SI{120}{\celsius}, indicating solute exchange with the far-field matrix \revision{was} tending to replace some of the solute lost to the GB precipitates.
Due to its higher stoichiometry in the GB precipitates and diffusivity, Zn exhibited a higher influx rate into the domain than Mg and had a\revision{n} \revision{average} composition of 8 at.\% \revision{for the whole domain} after \SI{54.2}{\hour} ageing.
However, the \revision{average} composition of Cu in the domain was almost constant during the entire ageing simulation, which reveals that the Cu in the matrix adjacent to the GB had a negligible interaction with the far-field matrix, due to its relatively low diffusivity.
The corresponding evolution of the composition profiles across the GB are shown in \cref{fig: Conc_profiles_fix_seg} (d - f).
Compared to the results using periodic boundary conditions (\cref{fig: Conc_profiles_Seg} (c)), it can be seen that Cu had a larger composition gradient and a lower average composition in the matrix close to the GB.  
However, Mg and Zn exhibited a higher composition level in the matrix close to the GB, resulting from the solute flux from the far-field matrix into the region close to the GB, especially for Zn which has high diffusivity.

\begin{figure}
	\centering
	\includegraphics[width=0.8\textwidth]{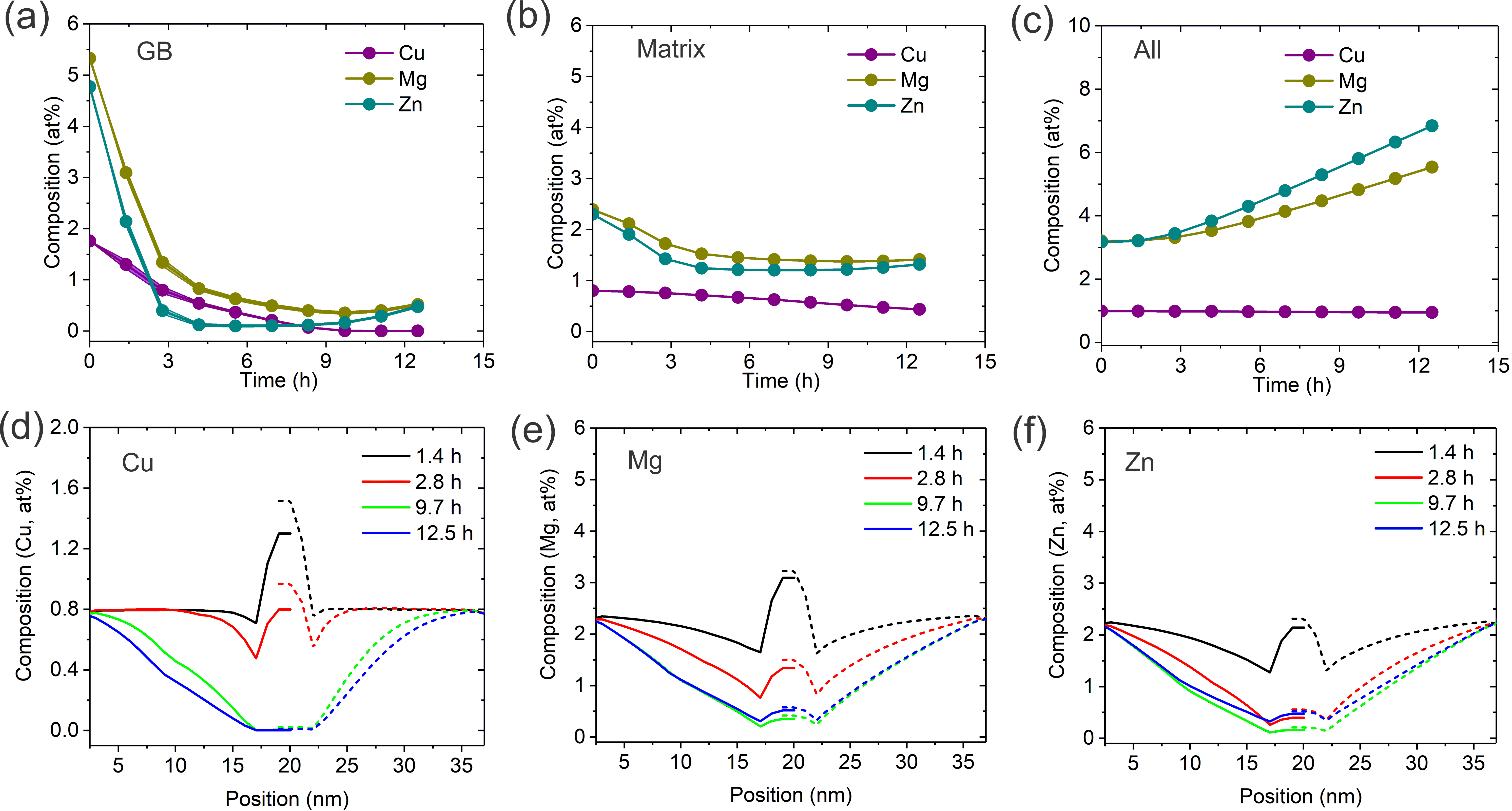}
	\caption{The effect of GB diffusion under constant potential  boundary conditions on the evolution of average solute composition (a) on the GB, (b) within the matrix adjacent to the GB, and (c) in the overall domain.
		Composition profiles of (d) Cu, (e) Mg, and (f) Zn across the GB with ageing at \SI{120}{\celsius}.
		The solid lines (left half) depict the average composition across the GB and the dashed lines (right half) are the corresponding results at a local point, as far \revision{away} as possible from a precipitate, indicated by the red arrow in \cref{fig: GB_setup}.
		Only half of the profiles are plotted due to the symmetry of the results.
		GB diffusion was included and assumed to be 4 to 5 orders of magnitude higher than the bulk diffusion.
		The chemical potentials of all the solute elements (Zn, Mg, Cu) were fixed on the boundaries of the simulation box, which allows the solute exchange between the matrix adjacent to the GB and the far-field matrix.
	}
	\label{fig: Conc_profiles_Seg_GBM_ConsFlux}
\end{figure}

\cref{fig: Conc_profiles_Seg_GBM_ConsFlux} shows the evolution of the solute distribution on the GB and within the region close to the GB, when both GB diffusion \revision{was} included and solute exchange \revision{was} allowed.
It can be seen that the solute flux from the far-field matrix significantly reduced GB Cu and Mg segregation, while Zn segregation was moderately enhanced (as shown in \cref{fig: Conc_profiles_Seg_GBM_ConsFlux} (a)).
In comparison, when periodic boundary conditions were used, rapid Zn exhaustion was the growth limiting process \revision{for the GB precipitates}, resulting in an excess \revision{of} Mg and Cu \revision{in the GB plane}.
However, with the influx of Zn from the far-field matrix, precipitate growth was no longer limited and this therefore allow\revision{ed} more consumption of the \revision{segregated} Mg and Cu.
In addition, in the region close to the GB, the Mg and Zn composition\revision{s} \revision{were} now substantially increased, while the composition of Cu was slightly decreased (\cref{fig: Conc_profiles_Seg_GBM_ConsFlux} (b)).
Similarly, there was an overall influx of Mg and Zn into the domain after 12.5h ageing, while the Cu level in the overall domain was almost constant  (\cref{fig: Conc_profiles_Seg_GBM_ConsFlux} (c)).
This indicates that, even in this extreme case, the width of the Cu depletion zone was still confined to be very local to the GB, and there was almost no interaction with the far-field matrix.
Consequently, the effect of the far-field matrix on accelerating GB precipitate growth \revision{was} mediated only through \revision{exchange of} Mg and Zn.
This result\revision{ed} in a lower Cu content in the matrix adjacent to the GB and larger Cu gradient across the GB (\cref{fig: Conc_profiles_Seg_GBM_ConsFlux} (d)) compared with the results obtained under the periodic boundary conditions (\cref{fig: Conc_profiles_Seg_GBM} (c)).

\subsubsection{Influence of the grain boundary precipitate nucleation rate}
\label{GB_nucleation}
It is generally recognised that the nucleation of GB $\eta$-precipitates can vary substantially on different GBs, \eg\ depending on the individual GB structure and misorientation \citep{unwin1969nucleation}.
The GB precipitate nucleation rate can thus have a major influence on the evolution of the GB microstructure during the subsequent ageing treatment.
In this section, a series of simulated heat treatments were performed at \SI{120}{\celsius} with GB precipitate number densities ranging from \SI{500}{\micro\meter^{-2}} to \SI{1500}{\micro\meter^{-2}}, to investigate the influence of the GB \revision{precipitate number density}.
This was controlled by systematically changing the number of precipitate seeds employed in the model. 

The influence of the number density of \revision{GB} precipitates on \revision{their} average \revision{transformation} kinetics is presented in \cref{fig: Volume_frac}, where the dashed lines and the dotted lines depict the results with lowest and highest precipitate number densit\revision{ies studied} (from \SI{500}{\micro\meter^{-2}} to \SI{1500}{\micro\meter^{-2}}).
It can be seen that, without including GB diffusion (comparing red lines), a higher precipitate number density resulted in a much larger overall precipitate \revision{transformation} rate at the beginning of ageing up to \SI{5}{\hour} at \SI{120}{\celsius}, but hereafter this enhancement effect decreased gradually with ageing time.
After \SI{54.2}{\hour} ageing at \SI{120}{\celsius}, a 3 times higher number density \revision{led} to an \SI{18.8}{\percent} relative increase of the total precipitate volume fraction.
However, the simulated results indicate that the inclusion of faster GB diffusion significantly diminished the influence of the number density of precipitates on the average precipitate \revision{transformation} kinetics (comparing purple lines).

\begin{figure}
	\centering
	\includegraphics[width=0.9\textwidth]{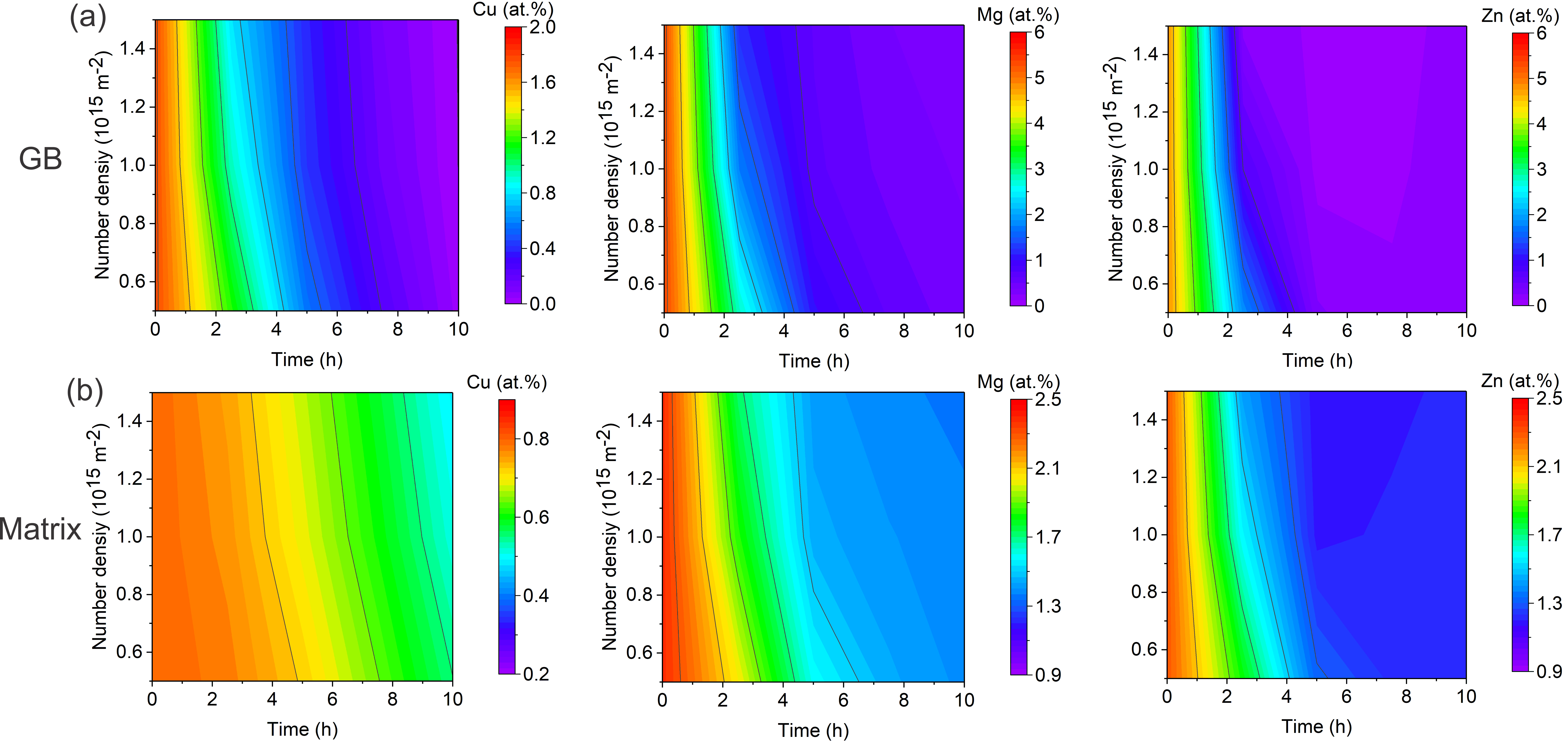}
	\caption{Influence of the GB precipitate number density on the evolution of (a) GB segregation, and (b) the compositions within the matrix close to the GB.
		GB diffusion was included and assumed to be 4 to 5 orders of magnitude higher than the bulk diffusion.
		Constant potential boundary condition was applied, where the chemical potentials of all the solute elements (Zn, Mg, Cu) were fixed on the boundaries of the simulation box, which allows the solute exchange between the matrix adjacent to the GB and the far-field matrix.
	}
	\label{fig: GB_nucleation_Fix_GBM02_rainbow}
\end{figure}

As shown in \cref{fig: GB_nucleation_Fix_GBM02_rainbow}, when fast GB diffusion \revision{was} included, the solute depletion rate both on the GB plane and within the matrix close to the GB was almost independent of the precipitate number density (within the range from \SI{500}{\micro\meter^{-2}} to \SI{1500}{\micro\meter^{-2}} studied in the current work). 
The iso-composition lines for the solute elements, however, shifted slightly towards a short\revision{er} ageing time with increasing number density of precipitates. 
It is also noteworthy that although Cu has a relatively low diffusivity in the matrix compared to Mg and Zn, fast GB diffusion significantly diminished the sensitivity of the Cu depletion rate to the GB precipitate density.
When GB diffusion was not considered, the solute depletion rate both on the GB plane and within the matrix close to the GB increased with increasing GB precipitate number density and Cu had a higher sensitivity to the GB precipitate number density compared \revision{to} Mg and Zn.

\section{Discussion}
\label{sec: discussion}
\subsection{Comparison and assessment of the simulations and experimental results}
\label{sec: comparison}
In the current work, both APT and STEM-EDS analysis show that gradual GB solute depletion occur\revision{red} in Al-Zn-Mg-Cu alloys during \revision{artificial} ageing treatment\revision{s} from the \revision{initial segregation seen after quenching}, but at different rates with respect to the main elements of interest, \eg\ the \revision{peak} Mg and Cu \revision{concentration on the GBs reduced} to about 3.0 at.\% and 1.5 at.\%, respectively. 
While all the simulation cases considered here were able to predict this general trend, the difference\revision{s} between the cases studied (\eg\ GB diffusion, solute flux from the far-field matrix, precipitate number density)  \revision{were} significant and provide\revision{d} insight into \revision{some of} the observed variability \revision{seen} in experimental results \citep{sha2011segregation,zhang2019dynamic,kairy2018role,zhao2018segregation}.

Periodic and fixed boundary conditions were used in the simulations, which physically represent extremely fine and coarse grain sizes, respectively.
The simulation results show that the final Mg and Cu GB \revision{segregation} after ageing was lower when fixed boundary conditions were used compared to with periodic boundary conditions, \ie\ it decreases with increasing grain size.
Under periodic boundary conditions, the Zn/Cu ratio in the GB precipitate (10.1) was much higher than that in the overall domain (3.2).
Consequently, with the alloy composition investigated\revision{,} which has an excess of Mg relative to the $\eta$-phase stoichiometry, Zn exhaustion was the precipitate growth-limiting process, leading to \revision{an excess in residual} GB Cu.
However, as shown in \cref{fig: Conc_profiles_Seg_GBM_ConsFlux} (c), under the fixed boundary condition, the Zn/Cu ratio in the domain increased from 3.2 to 7.2 after \SI{12.5}{\hour} ageing, owing to the net influx of Zn into the domain to supply the growing GB precipitate\revision{.} 
\revision{This occurred because of the Zn's} higher diffusivity compared to Cu, which did not interact with the far-field matrix.
\revision{Allowing the possibility of supplying Zn into the domain} thus result\revision{ed} in the continuous growth of the GB precipitates and consumption of the \revision{excess} GB Cu.
Analogously, \revision{the} greater influx of Zn than Mg \revision{into the domain} resulted in an increase in the Zn/Mg ratio from 0.99 to 1.2, which was still smaller than the GB precipitate ratio of 1.4, and will consequently result in the consumption of the GB \revision{excess} Mg.
These predictions are supported by similar observations of a decrease in Mg and Cu GB \revision{segregation} with increasing grain size found from APT experiments on an ultrafine grained AA7136 alloy, processed by equal-channel angular pressing \citep{sha2011segregation}, and  AA7075, processed by high pressure torsion \citep{zhang2019dynamic}.

\revision{In AA7050 alloy} STEM-EDS measured Mg and Cu residual segregation levels were found to be relatively constant along a GB after \revision{T7651} overageing (\cref{fig: STEM}), with no observable Zn excess, while the APT analysis, in \cref{fig: APT_compositions} (e), showed a slightly enriched Zn GB composition of 1.2 at.\%.
This variability in GB Zn composition can be partly explained by the different sensitivities of the two experimental techniques, but could also be caused by the effect of the GB diffusion on the evolution of the GB solute composition. 
As shown in \cref{fig: Conc_profiles_Seg} (a) and \cref{fig: Conc_profiles_Seg_GBM} (a), GB diffusion \revision{was} able to significantly reduce the GB solute excess at a given ageing time, with Zn being the most sensitive followed by Mg, whereas Cu \revision{was} relatively unaffected.
Due to the\revision{ir} higher GB diffusion rate, HAGBs acting as solute collectors provide a fast path to transport solute to the precipitates, resulting in the more rapid consumption of the GB solute excess and a homogeneous distribution within the GB plane.
As the solute mobility can vary significantly depending on the structural state of individual GBs, it can be postulated that a range of Zn GB \revision{segregation levels} may exist on different GBs, \revision{which could contribute to} the observed discrepancy between the APT and STEM-EDS analysis, however, more experimental data is required to confirm this predicted effect.
The simulations also show a larger variability in Mg GB \revision{segregation level}, with less variability in Cu.
These predictions are further supported by APT observations of a larger composition variation of GB Mg (0.9 at.\%) compared to Cu (0.3 at.\%) in \revision{an ultrafine grained} AA7136 alloy\citep{sha2011segregation}.

\subsection{Role of grain boundary segregation and diffusion in corrosion}
\label{sec: GBsegDiffusion}
The evolution of the GB microchemistry and microstructure is important in understanding the corrosion behaviour of Al-Zn-Mg-Cu alloys.
It is generally accepted that GB Mg segregation is detrimental to SCC resistance\revision{,} due to \revision{a} strong Mg-H interaction \citep{song2004stress}, while Cu enrichment on GBs and within GB precipitates can retard the anodic dissolution \revision{of GB $\eta$-phase precipitates} and enhance GB cohesion \citep{ramgopal2001electrochemical,knight2015some}.
It has also been reported that the GB segregation of Zn may be responsible for hydrogen embrittlement and the high susceptibility to SCC of Al-Zn-Mg-Cu alloys \citep{gruhl1984stress}.
However, this has since been disputed by subsequent research that showed no correlation between Zn segregation and SCC \citep{goswami2013evolution,knight2010correlations}.
In the current work, phase-field simulations as well as APT and STEM-EDS characterisation \revision{have} reveal\revision{ed} that Cu and Mg moderately segregated \revision{to} GB\revision{s} in peak and overaged conditions, \revision{while there was} little to no Zn \revision{residual} segregation\revision{, as any initially segregated Zn was rapidly consumed by the growing precipitates}.
This suggests that after overageing the SCC resistance in AA7050 alloys is predominantly affected by the Cu and Mg segregation, with Zn playing a more minor role.

The solute interaction terms used in the CALPHAD model, for AA7050 alloy at \SI{120}{\celsius}, are only around \SI{10.2}{\percent} of the total Gibbs energy, which implies a limited solute interaction \revision{between the solute species} on GBs.
Therefore, the observed GB segregation \revision{was} strongly influenced by the multi-component diffusion between the GB and the matrix close to the GB and the rate of removal of solute by the growing GBPs, with \revision{only} subtle co-segregation effects among different solutes.
As described in \cref{eq:diffusion-equation2}, \revision{the diffusion of component $i$ depends on} the chemical potential gradient\revision{s} \revision{of all the components, which are correlated by} the mobility matrix.
For AA7050 at \SI{120}{\celsius}, the mobilities for Cu in the matrix are $L_\text{CuCu}^\text{FCC}=$ \SI{5e-28}{\metre \squared \mole \joule^{-1} \second^{-1}}, $L_\text{CuMg}^\text{FCC}=$ \SI{-5.1e-28}{\metre \squared \mole \joule^{-1} \second^{-1}}, $L_\text{CuZn}^\text{FCC}=$ \SI{-5.8e-28}{\metre \squared \mole \joule^{-1} \second^{-1}}, indicating that Cu transport is significantly affected by uphill diffusion along gradients in the Mg and Zn composition.
In addition, the cross-mobilities for Mg and Zn, $L_\text{MgZn}^\text{FCC}=$ \SI{2.9e-27}{\metre \squared \mole \joule^{-1} \second^{-1}}, are approximately an order of magnitude smaller than their self-mobilities, $L_\text{MgMg}^\text{FCC}=$ \SI{5e-26}{\metre \squared \mole \joule^{-1} \second^{-1}} and $L_\text{ZnZn}^\text{FCC}=$ \SI{5.6e-26}{\metre \squared \mole \joule^{-1} \second^{-1}}, indicating that \revision{the effect of} Mg and Zn inter-diffusion is limited.
Since the magnitude of the Cu mobility is relatively small at \SI{120}{\celsius}, with $L_\text{ZnZn}^{FCC}/L_\text{CuCu}^{FCC}\approx$ 113, the large influx of Mg and Zn from the far-field matrix into the GB region \revision{thus} resulted in a minor outflux of Cu to the neighbouring bulk, as shown in \cref{fig: Conc_profiles_fix_seg} (c) and \cref{fig: Conc_profiles_Seg_GBM_ConsFlux} (c).
However, the GB self-mobility of Cu, $L_\text{CuCu}^\text{GB}=$ \SI{1.1e-22}{\metre \squared \mole \joule^{-1} \second^{-1}}, is about 13 times the GB cross-mobilities, $L_\text{CuMg}^\text{GB}=$ \SI{-8.1e-24}{\metre \squared \mole \joule^{-1} \second^{-1}} and $L_\text{CuZn}^\text{GB}=$ \SI{-8.8e-24}{\metre \squared \mole \joule^{-1} \second^{-1}}, indicating that interdiffusion of Cu on the GB is limited.
The rate of diffusion of Cu to the $\eta$-precipitates has important consequences on SCC resistance, as it was reported that Cu enrichment of $\eta$-precipitates reduces their anodic dissolution rate \citep{ramgopal2001electrochemical,knight2015some,knight2010correlations}.
Uphill diffusion is also a potential mechanism for Cu leaching from $\eta$-precipitates, which would be detrimental to SCC resistance.
The results show that uphill diffusion of Cu is active in the bulk at \SI{120}{\celsius}, but is minimal on the GB \revision{plane}.
Increasing the ageing temperature to \SI{180}{\celsius} is therefore expected to be beneficial for SCC resistance, as it not only further minimizes the interdiffusion of Cu (\ie\ reducing uphill diffusion of Cu) but also increases the mobility of Cu relative to Mg and Zn from $L_\text{ZnZn}^{FCC}/L_\text{CuCu}^{FCC}\approx$ 113 at \SI{120}{\celsius} to $L_\text{ZnZn}^{FCC}/L_\text{CuCu}^{FCC}\approx$ 6 at \SI{180}{\celsius}.
\revision{It also increases the equilibrium solubility of Cu in the $\eta$-phase which was not accounted for in the model \citep{marlaud2010influence,marlaud2010evolution}. }
Therefore, higher ageing temperature\revision{s} can increase the Cu/Zn ratio in the matrix adjacent to \revision{a} GB, which can enhance the enrichment of Cu along GBs and consequently within GB precipitates.
Improvement of the SCC resistance has been observed through high temperature stages in multi-stage \citep{wang2018two} and RRA \citep{knight2010correlations,gupta2012relating, knight2015some} heat treatments in Al-Zn-Mg-Cu alloys, in agreement with this prediction.

\begin{figure}
	\centering
	\includegraphics[width=0.9\textwidth]{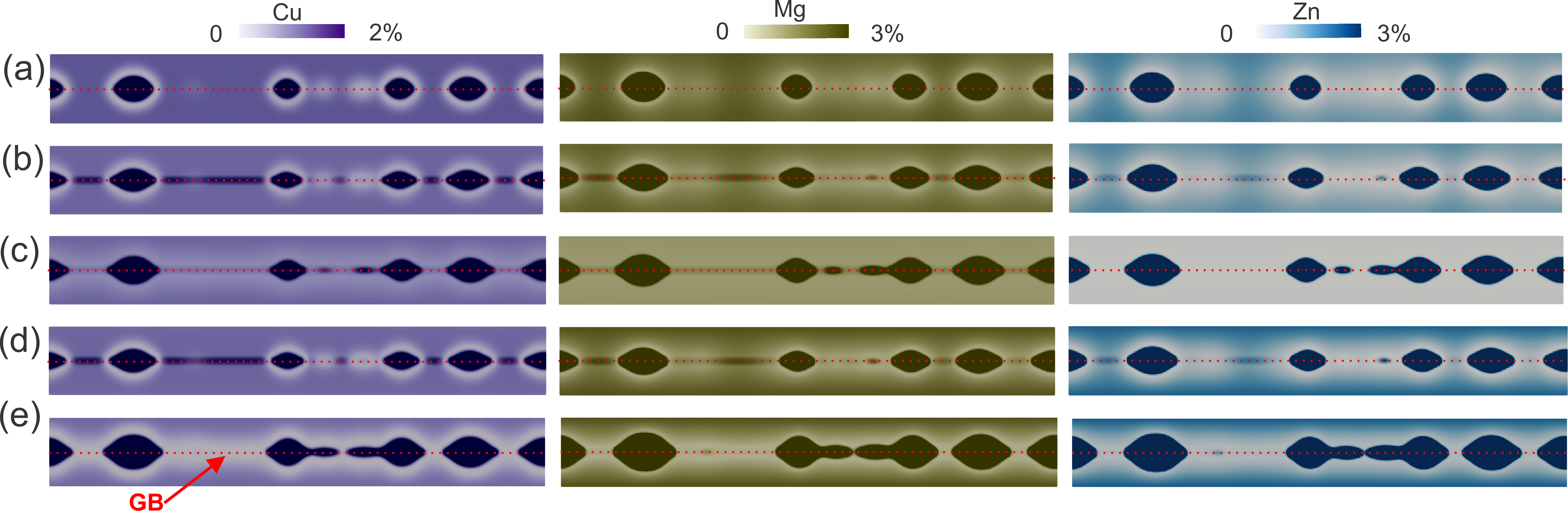}
	\caption{Effect of GB segregation, diffusion and boundary conditions on the solute distribution of Cu, Mg and Zn on the GB plane and within the matrix adjacent to the GB. 
		Shown is the cross section perpendicular to the GB plane, ageing after \SI{12.5}{\hour} at \SI{120}{\celsius}.
		(a) shows the results with periodic boundary conditions, without GB diffusion, and  without GB segregation;
		(b) shows the results with periodic boundary conditions, without GB diffusion, and with  GB segregation;
		(c) shows the results with periodic boundary conditions, with GB diffusion, and  with GB segregation;
		(d) shows the results with constant potential boundary conditions, without GB diffusion, and with GB segregation;
		(e) shows the results with constant potential boundary conditions, with GB diffusion, and with GB segregation.
	}
	\label{fig: GB_section}
\end{figure}

The simulation results also indicate that GB diffusion ha\revision{d} a large influence on the precipitate growth, with GB energetics having a minor effect.  
On increasing GB diffusivity, the solute depletion zones of Mg and Cu increase\revision{d} in size and distribute\revision{d} more homogeneously across \revision{a} GB, as shown in \cref{fig: GB_section}, indicating that GB diffusion can markedly accelerate the solute redistribution kinetics within the matrix adjacent to the GB.
With enhanced GB diffusion solute atoms from the matrix adjacent to the GB are collected onto the GB across a larger area through bulk diffusion and then diffuse rapidly along the GB to the precipitates by GB diffusion.
Since solute mobility on the GB can be 4 to 5 orders of magnitude larger compared to \revision{in} the bulk, for only a \SI{20}{\percent} reduction of the solute migration energy, this effect is \revision{very} important and GBs acting as solute collectors can effectively transport solutes across large distances, resulting in the accelerated growth of GB precipitates both along\revision{,} as well as normal to\revision{,} the GB.
This \revision{effect} can thus explain the observation of the GB precipitates with much larger thickness than those nucleated in the matrix at the same time, as characterised by APT (\cref{fig: APT_ppts} (d)) and TEM experiments (\cref{fig: STEM} (a)).

Finally, it should be noted that in these simulations to simplify the model a fixed stoichiometry was adopted for the $\eta$-phase, although this is unrealistic. 
For example, it is known that in particular the Cu content of the GB precipitates increases with ageing time, owing to its lower diffusivity, and the equilibrium $\eta$-phase composition also changes with ageing temperature \citep{marlaud2010influence,marlaud2010evolution}. 
This effect might be expected to increase the level of Cu on the GB plane at short ageing time\revision{s} but should still converge with our simulations at longer ageing times\revision{,} as we used the composition of the $\eta$-phase measured when it was close to equilibrium. 
In addition, nucleation of precipitates in the matrix was ignored. 
Matrix precipitation will act in competition with the GB precipitates for solute, and therefore inhibit their growth at longer ageing times; although to some extent this was replicated by the predictions \revision{performed} with periodic boundary conditions. 
\revision{Furthermore, the formation possibility of quench-induced GB precipitation occurring during cooling after solution treatment was not considered in the current phase-field simulations.
Owing to the quench sensitivity of Al-Zn-Mg-Cu alloys and the low cooling rate experienced in thick plates during quenching, large quench-induced precipitates (\textit{e.g.} $\eta$-phase, S-phase, T-phase) can also occur on GBs, prior to those that subsequently nucleate during ageing \citep{godard2002precipitation,GARNER2021190}.
These large quench-induced GB precipitates, being of the order of around \SI{1}{\micro \metre}, can substantially consume the solutes along GBs and in the matrix adjacent to GBs, which will play an important role in the formation and compositional evolution of the subsequent age-induced GB precipitates and the formation of their associated precipitate free zones.} 

\section{Conclusions}
\label{sec: conclusions}
In this work, a \revision{model involving} direct coupling of the phase-field method \revision{with} CALPHAD \revision{thermodynamic phase descriptions} has been developed to predict complex and transient microstructure transformations in multi-component engineering alloys, and an efficient numerical implementation has been proposed to solve the resulting evolution equations.

The developed model has been applied to simulation of the growth of a population of GB $\eta$-precipitates in \revision{an} Al-Zn-Mg-Cu alloy as an engineering case study.
This work \revision{has} systematically investigated the influence of the GB solute segregation, GB diffusion, the far-field matrix and the GB precipitate nucleation rate on the GB precipitation and microchemistry evolution.
In agreement with APT and TEM observations, significant Mg and Cu GB segregation was predicted to remain after peak ageing, while Zn \revision{was} depleted at an early stage in the \revision{ageing} heat treatment.
Solute segregation and the respective very different mobilities of the species significantly affected the solute partitioning on GBs and the morphological evolution of GB precipitates, however, it had little influence on the average precipitate \revision{transformation} kinetics.
The simulation results reveal\revision{ed} that fast GB diffusion accelerate\revision{d} precipitate growth by \revision{more} effectively transporting solutes from far-field regions to the solute depletion zones surrounding the precipitates.
Cu diffusion \revision{was} observed to be confined within the matrix adjacent to the GB due to its lower concentration in the precipitate\revision{s} and low diffusivity, whereas the solute depletion zones of Mg and Zn reached beyond the matrix adjacent to the GB at an early stage on ageing at \SI{120}{\celsius}.
Finally, it \revision{was} found that the influence of precipitate nucleation on the  local heterogeneity of the chemical composition evolution on the GB plane, and within the adjacent regions, can be significantly diminished by fast GB diffusion pathways.

\section{Acknowledgements}

PS and PBP are grateful to the EPSRC for financial support through the associated programme grant LightFORM (EP/R001715/1) and the Airbus\textendash University of Manchester Centre for Metallurgical Excellence, UK for supporting aspects of this research.
PS and CL are grateful to the DFG for financial support through subproject M5 in the Priority Programme SPP 1713: Strong Coupling of Thermo-chemical and Thermo-mechanical States in Applied Materials.
\revision{CL also acknowledges the kind support by Shanghai Jiao Tong University through the Outstanding Graduate Student program.}

\section{References}
\bibliographystyle{elsarticle-num}
\bibliography{\jobname}

\appendix 
\section{\revision{Influence of the misfit strain on $\eta$-precipitate growth}}
\label{sec: elastic}
\revision{
A three-dimensional single crystal setup of \SI{60}{\nano \metre} $\times$ \SI{60}{\nano \metre} $\times$ \SI{60}{\nano \metre} with an isolated $\eta$-precipitate was used here.
An initial spherical precipitate with a radius of \SI{2.5}{\nano \metre} was located at the centre of the cubic box. 
The homogeneous modulus approximation was employed in the simulation, \ie\ both the matrix and the precipitate were assumed to have the same stiffness tensor \stiffness\ ($C_\text{11} = \text{106 GPa}$, $C_\text{12} = \text{60 GPa}$, $C_\text{44} = \text{28 GPa}$) \citep{vallin1964elastic}.
The partial rank-one mechanical homogenisation approach has been applied for the stress calculation of the interface region, which satisfies the mechanical jump conditions \citep{schneider2017stress}.
The other thermodynamic and kinetic material parameters used are the same as described in \cref{sec: validation,sec: setup}. 
Two different sets of stress-free transformation strains were applied to the $\eta$-precipitate, \ie\ 
}

\revision{
	$\mathbf{F}_\text{co} = \begin{bmatrix}
	0.95 & 0 & 0\\
	0 & 1 & 0\\
	0.15 & 0 & 1
\end{bmatrix}$
and
$\mathbf{F}_\text{in} = \begin{bmatrix}
	0.99 & 0 & 0\\
	0 & 1 & 0\\
	0.03 & 0 & 1
\end{bmatrix}$.
}

\revision{$\mathbf{F}_\text{co}$ and $\mathbf{F}_\text{in}$ are the transformation deformation gradient tensors of the coherent and incoherent precipitate, respectively.}

\begin{figure}
	\centering
	\includegraphics[width=0.75\textwidth]{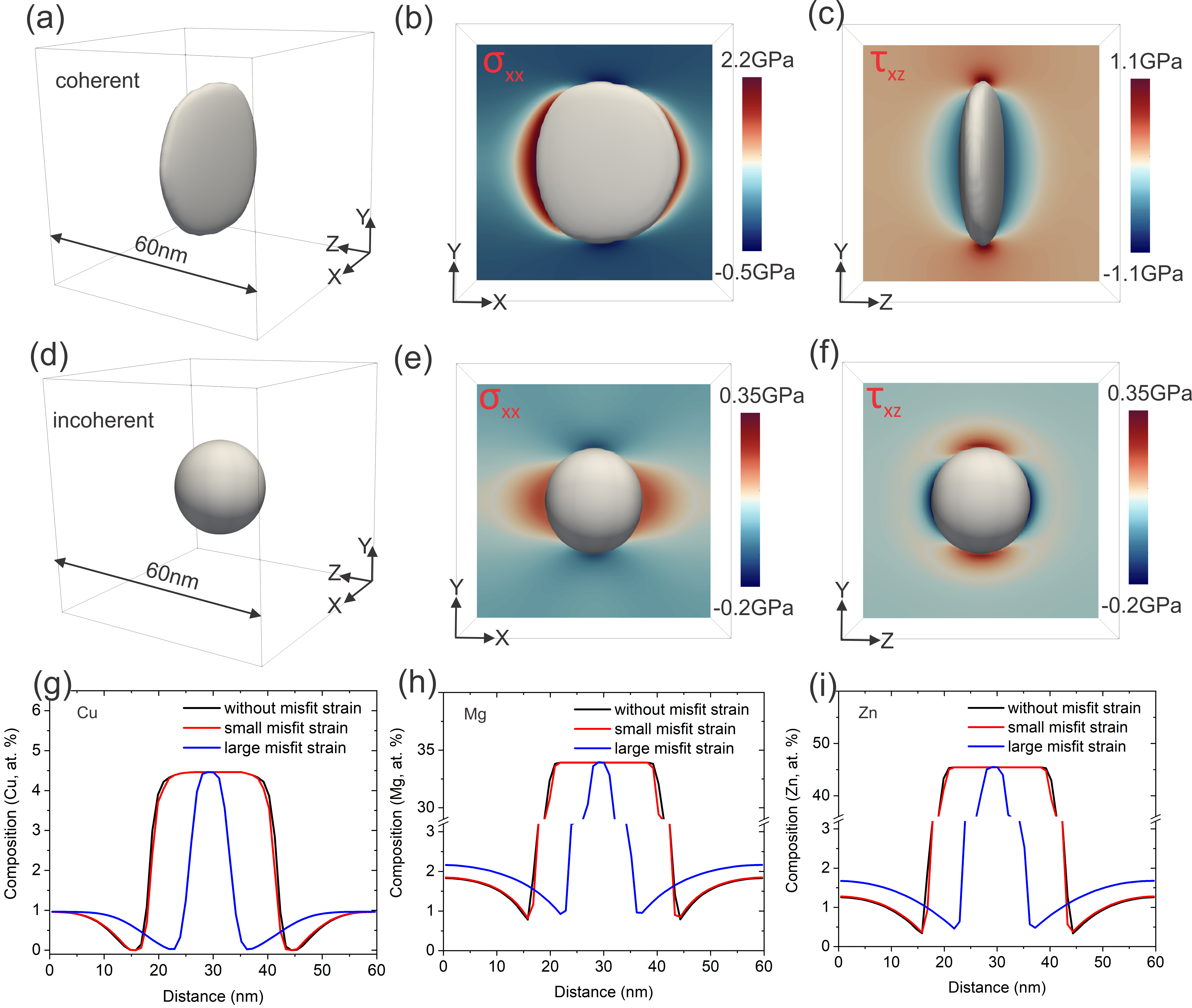}
	\caption{\revision{ The phase-field simulated shape of the $\eta$-precipitate aged after \SI{55}{\hour} at \SI{120}{\celsius} for (a) coherent and (d) incoherent cases, respectively.
		The corresponding contour plots of the calculated (b, e) nominal ($\sigma_{\text{xx}}$)  and (c, f) shear stress fields ($\tau_{\text{xz}}$) in the matrix induced by the transformation strain of the precipitate.
		The corresponding compositional profiles of (g) Cu, (h) Mg, and (i) Zn across the centre of the single crystal alone Z direction.} }
	\label{fig: Elas_Cauchy}
\end{figure}

\revision{
\cref{fig: Elas_Cauchy} shows the simulated shape of the $\eta$-precipitate and the stress fields in the matrix after ageing for \SI{55}{\hour} at \SI{120}{\celsius}. 
It can be seen that the coherent precipitate had an plate shape (\cref{fig: Elas_Cauchy} (a)), which is dominated by the minimization of the elastic strain energy.
As shown in \cref{fig: Elas_Cauchy} (c), a high negative magnitude of shear stress can be observed in the matrix around the broad face of the coherent precipitate, due to the large shear Eigen strain along the normal of the broad face.
Since the misfit strain for the incoherent precipitate is relatively small (a combination of 3\% shear and 1\% compression deformation), the precipitate still exhibited a spherical shape (\cref{fig: Elas_Cauchy} (d)) after ageing for \SI{55}{\hour} at \SI{120}{\celsius}, where the anisotropic elastic strain energy played a minor role in this case.
\cref{fig: Elas_Cauchy} (g - i) show the corresponding compositional profiles across the centre of the single crystal along Z direction.
It is clearly shown that all the solute profiles (Cu, Mg, Zn) in the case of incoherent precipitate almost overlapped with these without any misfit strain.
Again, it demonstrates that the isotropic chemical free energy and interface energy, comparing with elastic strain energy, dominated the incoherent precipitate growth behaviour.
Therefore, these results imply that it is physically reasonable to neglect the elastic strain energy contribution for simulation of the GB incoherent precipitation behaviour in Al-Zn-Mg-Cu alloys. 
}

\section{\revision{Influence of the interface energy on grain boundary precipitation}}
\label{sec: interface}
\revision{
A three-dimensional bi-crystal setup of \SI{40}{\nano \metre} $\times$ \SI{40}{\nano \metre} $\times$ \SI{40}{\nano \metre} with an isolated GB $\eta$-precipitate was employed here.
An initial spherical precipitate with a radius of \SI{2.5}{\nano \metre} was located at the centre of the GB plane.
The thermodynamic and kinetic material parameters used are the same as described in \cref{sec: validation,sec: setup}.
Since the incoherent GB $\eta$-phases generally do not have specific orientation relation with the matrix \citep{unwin1969nucleation,butler1976situ,liu2010revisiting}, isotropic interface energy for the precipitate-matrix interface was applied here.
Furthermore, the precipitate-matrix interface energy and the GB interface energy were assumed to be same.
In order to reveal the effect of interface energy on the GB segregation and precipitation evolution behaviour, three phase-field simulations with varying interface energies of \SI{0.2}{\joule\metre^{-2}}, \SI{0.3}{\joule\metre^{-2}}, and \SI{1.0}{\joule\metre^{-2}} have been performed.
}

\begin{figure}
	\centering
	\includegraphics[width=0.7\textwidth]{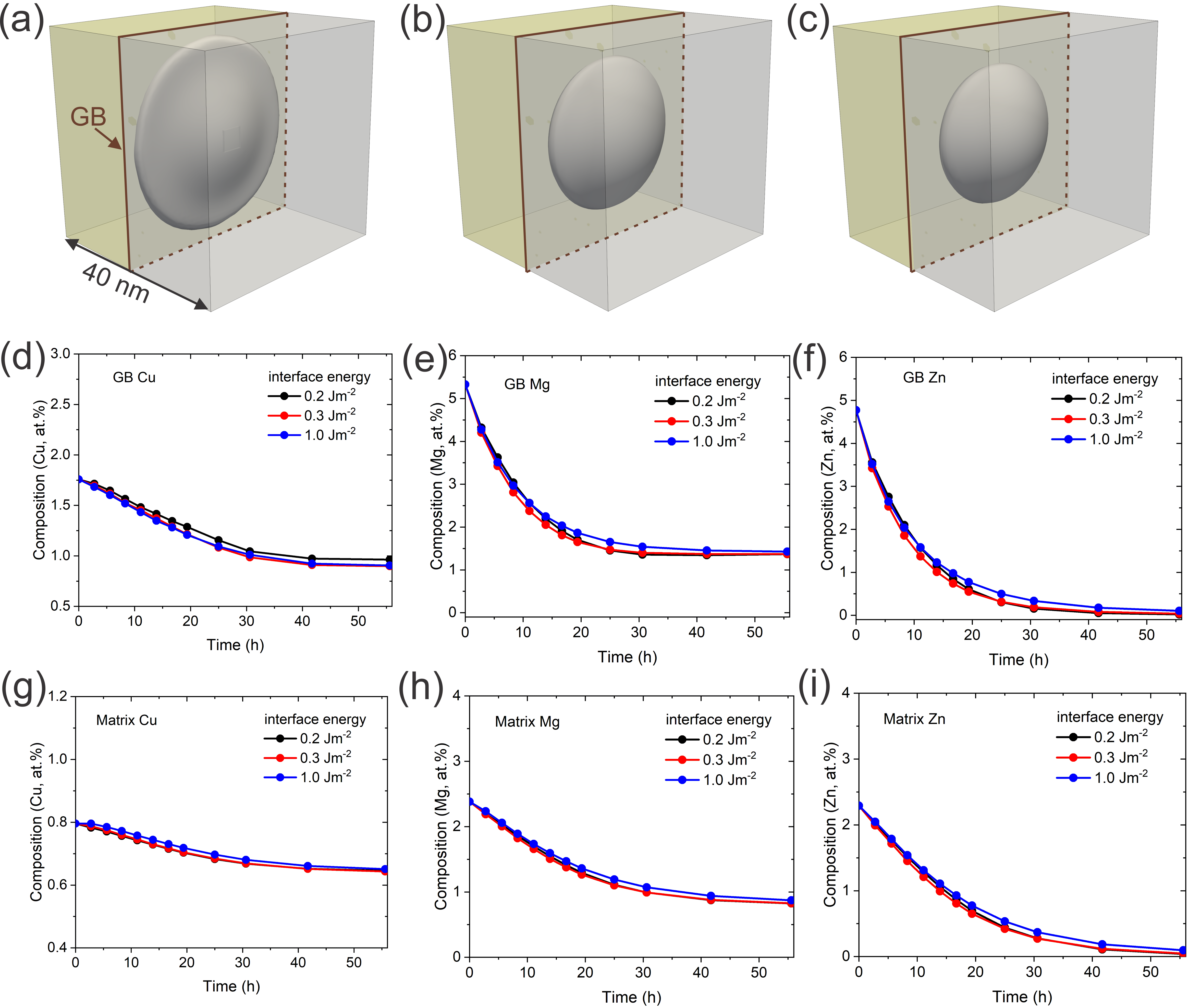}
	\caption{\revision{
			The phase-field simulated shape of the GB $\eta$-precipitate aged after \SI{55}{\hour} at \SI{120}{\celsius} with different interface energies of (a) \SI{0.2}{\joule\metre^{-2}}, (b) \SI{0.3}{\joule\metre^{-2}}, and (c) \SI{1.0}{\joule\metre^{-2}}.
		    Evolution of the simulated average solute compositions of Cu, Mg, and Zn; (d, e, f) on the GB without precipitates, and (g, h, i) within the matrix adjacent to the GB.
		    }
	}
	\label{fig: interface}
\end{figure}

\revision{
\cref{fig: interface} (a - c) show the corresponding simulated shape of the GB $\eta$-precipitate with different interface energies, after ageing for \SI{55}{\hour} at \SI{120}{\celsius}.
It can be seen that the shape of the GB $\eta$-precipitate was dominated by the interplay between surface tension forces at the GB-precipitate junction and solute segregation on the GB plane.
With increasing interface energy, the GB $\eta$-precipitate evolved from a plate shape to an ellipsoid shape, which agrees with the Wulff plot \citep{burton1951growth}.
\cref{fig: interface} (d - i) show the evolution of the average compositions on the GB without precipitates, and within the matrix adjacent to the GB, as a function of ageing time, respectively.
Almost the same results have been obtained with different interface energies, which revealed that the evolution of the GB segregation level and solute depletion within the matrix was almost independent on the interface energy .
These results indicated that interface energy played a minor role in the compositional evolution of the GB region (within the range from \SI{0.2}{\joule\metre^{-2}} to \SI{1.0}{\joule\metre^{-2}} studied in the current work).
}

\section{Thermodynamic and kinetic material parameters}
\label{sec: params}

\begin{table}[h!]
	\centering
	\caption{Thermodynamic data for the Al-Cu-Mg-Zn system (J/mol)}
	\begin{tabular}{llll}
		\toprule
		FCC: Disordered solution\tabularnewline
		\midrule
		$~^0L_{\text{AL,Cu}}^\text{fcc} = -53520 + 2\text{T}$\tabularnewline
		$~^1L_{\text{AL,Cu}}^\text{fcc} = 38590 - 2\text{T}$\tabularnewline
		$~^2L_{\text{AL,Cu}}^\text{fcc} = 1170$\tabularnewline
		$~^0L_{\text{AL,Mg}}^\text{fcc} = 4971 -3.5\text{T}$\tabularnewline
		$~^1L_{\text{AL,Mg}}^\text{fcc} = 900 +0.423\text{T}$\tabularnewline
		$~^2L_{\text{AL,Mg}}^\text{fcc} = 950$\tabularnewline
		$~^0L_{\text{AL,Zn}}^\text{fcc} = 7297+0.475\text{T}$\tabularnewline
		$~^1L_{\text{AL,Zn}}^\text{fcc} = 6613 -4.59\text{T}$\tabularnewline
		$~^2L_{\text{AL,Zn}}^\text{fcc} = -3097 + 3.3\text{T}$\tabularnewline
		$~^0L_{\text{Cu,Mg}}^\text{fcc} = -22279.28 + 5.868\text{T}$\tabularnewline
		$~^0L_{\text{Cu,Zn}}^\text{fcc} = -42803.75+10.02\text{T}$\tabularnewline
		$~^1L_{\text{Cu,Zn}}^\text{fcc} = 2936.39 -3.05\text{T}$\tabularnewline
		$~^2L_{\text{Cu,Zn}}^\text{fcc} = 9034.2 -5.39\text{T}$\tabularnewline
		$~^0L_{\text{Mg,Zn}}^\text{fcc} = -3056.82+5.64\text{T}$\tabularnewline
		$~^1L_{\text{Mg,Zn}}^\text{fcc} = -3127.26 +5.66\text{T}$\tabularnewline
		$~^1L_{\text{AL,Cu,Mg}}^\text{fcc} = 60000$\tabularnewline
		$~^1L_{\text{AL,Cu,Zn}}^\text{fcc} = 72990.79$\tabularnewline
		$~^2L_{\text{AL,Cu,Zn}}^\text{fcc} = 16799.93$\tabularnewline
		$~^3L_{\text{AL,Cu,Zn}}^\text{fcc} = 59580.95$\tabularnewline
		\midrule
		Stochiometric $\eta$ phase: $(\text{Zn}_{45.4}, \text{Al}_{16.8}, \text{Cu}_{4.5} )\text{Mg}_{33.3}$\tabularnewline
		\midrule
		$G_{(\text{Zn}_{45.4}, \text{Al}_{16.8}, \text{Cu}_{4.5} )\text{Mg}_{33.3}} = -26605 $ at 393K\tabularnewline
		\bottomrule
	\end{tabular}
	\label{Tab: Thermodynamic}
\end{table}

\begin{table}[h!]
	\centering
	\caption{Solute mobility parameters for the Al-Cu-Mg-Zn system in the FCC phase: Migration energy (J/mol)}
	\begin{tabular}{llll}
		\toprule
		Mobility of Al\tabularnewline
		\midrule
		$Q_\text{Al}^\text{Al} = -127200 -92.9858\text{T}$\tabularnewline
		$Q_\text{Al}^\text{Cu} = -181583.4 -99.8\text{T}$\tabularnewline
		$Q_\text{Al}^\text{Mg} = -127200 -92.9858\text{T}$\tabularnewline
		$Q_\text{Al}^\text{Zn} = -83255 -92.9262\text{T}$\tabularnewline
		$~^0 Q_\text{Al}^\text{Al,Zn} = 30169 -111.8367\text{T}$\tabularnewline
		$~^1 Q_\text{Al}^\text{Al,Zn} = 11835 +39.0022\text{T}$\tabularnewline
		\midrule
		Mobility of Cu\tabularnewline
		\midrule
		$Q_\text{Cu}^\text{Al} = -131000 -81.3999\text{T}$\tabularnewline
		$Q_\text{Cu}^\text{Cu} = -204670 -83.1464 \text{T}$\tabularnewline
		$Q_\text{Cu}^\text{Mg} = -112499 -81.26\text{T}$\tabularnewline
		$Q_\text{Cu}^\text{Zn} = -91608.6 -82.63\text{T}$\tabularnewline
		$~^0 Q_\text{Cu}^\text{Al,Cu} = -31461.4 +78.91\text{T}$\tabularnewline
		$~^0 Q_\text{Cu}^\text{Al,Mg} = 175000$\tabularnewline
		$~^0 Q_\text{Cu}^\text{Al,Zn} = -250000$\tabularnewline
		$~^0 Q_\text{Cu}^\text{Cu,Zn} = -4930.8 -24.75\text{T}$\tabularnewline
		\midrule
		Mobility of Mg\tabularnewline
		\midrule
		$Q_\text{Mg}^\text{Al} = -119000 -95.7238\text{T}$\tabularnewline
		$Q_\text{Mg}^\text{Cu} = -170567 -98.84 \text{T}$\tabularnewline
		$Q_\text{Mg}^\text{Mg} = -112499 -81.2527\text{T}$\tabularnewline
		$Q_\text{Mg}^\text{Zn} = -71147 -67.009\text{T}$\tabularnewline
		$~^0 Q_\text{Mg}^\text{Al,Cu} = 200000$\tabularnewline
		$~^0 Q_\text{Mg}^\text{Al,Mg} = 53551$\tabularnewline
		$~^0 Q_\text{Mg}^\text{Cu,Mg} = 125000$\tabularnewline
		$~^1 Q_\text{Mg}^\text{Cu,Mg} = 50000$\tabularnewline
		$~^2 Q_\text{Mg}^\text{Cu,Mg} = 50000$\tabularnewline
		\midrule
		Mobility of Zn\tabularnewline
		\midrule
		$Q_\text{Zn}^\text{Al} = -120000 -88.4447\text{T}$\tabularnewline
		$Q_\text{Zn}^\text{Cu} = -190832.4 -86.1 \text{T}$\tabularnewline
		$Q_\text{Zn}^\text{Mg} = -73706 -86.1574\text{T}$\tabularnewline
		$Q_\text{Zn}^\text{Zn} = -76569 -86.2102\text{T}$\tabularnewline
		$~^0 Q_\text{Zn}^\text{Al,Zn} = -40720 + 31.7\text{T}$\tabularnewline
		$~^1 Q_\text{Zn}^\text{Al,Zn} = 147763 - 133.7\text{T}$\tabularnewline
		$~^0 Q_\text{Zn}^\text{Cu,Zn} = -68455.6 +36\text{T}$\tabularnewline
		\bottomrule
	\end{tabular}
	\label{Tab: kinetic}
\end{table}

\end{document}